\documentclass[11pt,a4paper]{article}
\pdfoutput=1
\usepackage{jheppub}
\usepackage{slashed}
\usepackage{comment}

\usepackage[english]{babel}

\usepackage{amsmath,bm}
\usepackage{amssymb}
\usepackage{graphicx}
\usepackage{ulem}
\usepackage{multirow}
\usepackage{xcolor,color}
\usepackage{booktabs}
\allowdisplaybreaks

\definecolor{frfgreen}{rgb}{0.16,0.67,0.47}

\definecolor{gesfblack}{rgb}{0,0,0}

\definecolor{gesfblue}{rgb}{0.08,0.42,0.76}

\definecolor{gesfgreen}{rgb}{0,1,0}

\definecolor{gesfgrey}{rgb}{0.5,0.5,0.5}

\definecolor{gesflanse}{rgb}{0.00,0.50,0.50}

\definecolor{gesfpurple}{rgb}{0.47,0.19,0.42}

\definecolor{gesfred}{rgb}{1,0,0}

\definecolor{gesfwhite}{rgb}{1,1,1}

\definecolor{gesfyellow}{rgb}{0.7,0.4,0.3}

\newcommand{\gsec}[1]{{\hypersetup{linkcolor=red}Sec.\,\ref{#1}\hypersetup{linkcolor=blue}}}
\newcommand{\gapp}[1]{{\hypersetup{linkcolor=red}Appendix~\ref{#1}\hypersetup{linkcolor=blue}}}
\newcommand{\geqn}[1]{\hypersetup{linkcolor=blue}Eq.\,(\ref{#1})\hypersetup{linkcolor=blue}}
\newcommand{\gfig}[1]{{\hypersetup{linkcolor=violet}Fig.\,\ref{#1}\hypersetup{linkcolor=blue}}}
\newcommand{\gtab}[1]{{\hypersetup{linkcolor=gesflanse}Table~\ref{#1}\hypersetup{linkcolor=blue}}}

\begin{document}

\title{Probing Light Dark Particles in Neutrino Scattering Experiments}

\author{}
\author{Ruofei Feng$^{1}$, Shao-Feng Ge$^{2,3}$, Yongchao Zhang$^{1,4}$}

\affiliation{$^1$School of Physics, Southeast University, Nanjing 211189, China}

\affiliation{$^2$State Key Laboratory of Dark Matter Physics, Tsung-Dao Lee Institute \& School of Physics and Astronomy, Shanghai Jiao Tong University, China}
\affiliation{$^3$Key Laboratory for Particle Astrophysics and Cosmology (MOE) \& Shanghai Key Laboratory for Particle Physics and Cosmology, Shanghai Jiao Tong University, Shanghai 200240, China}
\affiliation{$^4$Center for High Energy Physics, Peking University, Beijing 100871, China}

\emailAdd{fengrf@seu.edu.cn, gesf@sjtu.edu.cn, zhangyongchao@seu.edu.cn}


\abstract{
In this work we investigate the production of a dark fermionic particle $\chi$ in the neutrino scattering experiments. In the framework of effective field theory, such process can be induced by the effective four-fermion interactions involving neutrinos, the dark particle $\chi$ and standard model particles. We perform a comprehensive analysis of all possible Lorentz structures, considering representative neutrino experiments with distinct neutrino sources and target particles. In particular, we examine the constraints on the effective couplings for the neutrino-nucleus scattering by the latest COHERENT CsI and CONUS+ data, 
as well as the prospects at the DUNE near detector from neutrino-electron scattering. It turns out the current COHERENT and CONUS+ constraints on the cutoff scales are less stringent than those from the existing Large Hadron Collider data and the SN1987A observations. However, the DUNE near detector could probe the cutoff scales beyond the existing CHARM II and LEP limits up to roughly 1\,TeV, for the dark particle mass up to roughly 50\,MeV. Our results demonstrate the complementarity between neutrino experiments and collider searches in probing the dark sector physics.

}


\maketitle


\section{Introduction}

With the very recent updates of neutrino oscillation data by the JUNO experiment~\cite{JUNO:2025gmd}, we have entered the precision neutrino physics era at the sub-percentage level. However, neutrinos are still the most mysterious particles in the standard model (SM). For instance, we do not know much about the  nature of neutrinos (Dirac or Majorana particles), the origin of their tiny masses, or whether there are extra heavier states, etc~\cite{SajjadAthar:2021prg}. On the other hand, after decades of efforts, dark matter (DM) has not been undoubtedly confirmed in any of the direct detection, indirect detection or high-energy collider searches, in spite of the various astrophysical and cosmological evidence
\cite{Bertone:2004pz,Bertone:2016nfn,Roszkowski:2017nbc,Cirelli:2024ssz}. Besides DM, there could exist a hidden sector of dark particles to be explored. Therefore, the new physics (NP) beyond the SM of particle physics is essential for understanding the basic properties of neutrinos and dark particles, as well as other open questions in the SM.


Although both neutrinos and the dark particle $\chi$ are always ``invisible'' particles in the laboratory experiments and astrophysical observations, they could potentially interact sizably with each other (and the SM particles); see Ref.~\cite{Dev:2025tdv} for a recent comprehensive study. In the framework of effective field theory (EFT), there might exist the interactions of neutrinos $\nu$ and dark fermion $\chi$ with SM quarks or charged leptons $f = q,\ell$, in the form of effective four-fermion absorption operators of 
\begin{equation}
\label{eqn:operator}
  {\cal O}_i
\equiv (\bar\chi \Gamma^i P_L \nu) (\bar{f}\Gamma_if),
\end{equation}
where $P_{L} \equiv (1 - \gamma^5)/2$ is the left-handed projector, and $\Gamma_i$ stands for different Lorentz structures. Such effective couplings can originate from the simplified models with heavy mediators~\cite{Chang:2020jwl,Chen:2021uuw}. These EFT operators can induce very rich phenomenological signals. Depending on the underlying processes involved, the phenomenological implications can be classified into the following categories (for the sake of cleanness, the charge conjugate processes are not explicitly shown, e.g. $f\bar{f} \to \nu \bar\chi$ implies also the process $f\bar{f} \to\chi \bar\nu$ by charge conjugation). 
\begin{itemize}
    \item The decay of $\chi$. The most immediate implication of the operator in 
\geqn{eqn:operator} is the decay of the dark fermion $\chi$ at the tree-level $\chi \to \nu f\bar{f}$ or at the 1-loop order $\chi \to \nu\gamma,\, \nu\gamma\gamma,\, \nu\gamma\gamma\gamma$~\cite{Dror:2020czw,Ge:2022ius}. To stabilize a decaying DM at the cosmological scale, there are some requirements on the mass $m_\chi$ and couplings relevant, depending on the decay channels.
    However, our current study does not rely on the dark fermion $\chi$ to be a DM candidate but takes it as a dark sector particle for generality.
    
    \item $f^{(\ast)} + \bar{f}^{(\ast)} \to \nu + \bar\chi$. At high-energy hadron colliders, $f$ is the parton quarks involved, and neutrinos and the dark particle $\chi$ can be pairly produced and behave as missing transverse energy $\slashed{E}_T$. The dominant search channels are the mono-jet, mono-$\gamma$ and mono-$Z/W$ processes~\cite{Dror:2019onn,Ma:2024tkt,Ma:2024gqj}, which are very similar to the signals of dark particles at hadron colliders~\cite{Kahlhoefer:2017dnp,Boveia:2018yeb,Krnjaic:2022ozp}. At lepton colliders, $f$ denotes the charged leptons, and the most promising  signals are mono-$\gamma$ and $e^+ e^- + \slashed{E}_T$ processes~\cite{Ge:2023wye}. There have been very extensive studies of dark particles at the lepton colliders~\cite{Profumo:2009tb,Fox:2011fx,Bartels:2012ex,Dreiner:2012xm,Chae:2012bq,Freitas:2014jla,Bertuzzo:2017lwt,Habermehl:2020njb,Bharadwaj:2020aal,Kalinowski:2021tyr,Kundu:2021cmo,Barman:2021hhg,Liang:2021kgw,Chu:2018qrm,Krnjaic:2022ozp,Voronchikhin:2023znz,Asai:2023dzs,Liang:2023lwh} and from (semi-)invisible meson decays~\cite{McElrath:2005bp,Fayet:2006sp,Fayet:2007ua,Dreiner:2009er,McKeen:2009rm,Yeghiyan:2009xc,Badin:2010uh,Fernandez:2014eja,Fernandez:2015klv,Bertuzzo:2017lwt,Darme:2020ral,Bertuzzo:2020rzo,Schuster:2021mlr,Li:2021phq,Zhevlakov:2023wel,Gninenko:2024ujp}. At the 1-loop order, the operator induces the exotic invisible decay mode $Z \to \nu \bar\chi$, with $f$ and $\bar f$ being virtual particles in the loop~\cite{Davidson:2003ha}.
    



    \item $f^{\ast} \to f^{(\ast)} + \nu + \bar\chi$. Here $f^\ast$ stands for off-shell SM particles, e.g. internal quark or charged lepton lines. For instance, the operators could induce the rare meson decay ${\sf M}^- \to \ell^- \nu \nu \bar\chi$ and the exotic $W$ boson decay $W \to \ell \nu \nu \bar\chi$. Such processes are quite similar to the neutrino self-interaction induced processes ${\sf M}^- \to \ell^- \nu \nu \bar\nu$ and $W \to \ell \nu \nu \bar\nu$~\cite{Bilenky:1999dn,Bardin:1970wq,Kelly:2019wow,Bilenky:1992xn}. If dark particle is involved, the pair production of dark particles in the fixed-target experiments are also of this category~\cite{Kahn:2014sra,Batell:2014mga,Chu:2018qrm,Berlin:2018bsc,Banerjee:2019pds,Dutta:2019nbn,Berlin:2020uwy,Blinov:2020epi,Bertuzzo:2020rzo,Liang:2021kgw,Schuster:2021mlr,Krnjaic:2022ozp,Costa:2022pxv,Zhevlakov:2022vio,Abdullahi:2023tyk,Gninenko:2023rbf}.     
    The couplings of neutrino and dark particle $\chi$ with up-type quarks will contribute at the 1-loop order to the flavor-changing neutral current (FCNC) processes $d_j \to d_i \nu \bar\chi$ (with $d_{i,\,j}$ the down-type quarks and $i,\,j$ the generation indices), which have the same experimental signals as the SM FCNC process $d_j \to d_i \nu \bar\nu$ (including the extra contribution from neutrino non-standard interactions (NSIs)~\cite{Chen:2007cn}) and the DM pair production in FCNC meson decay $d_j \to d_i \chi \bar\chi$~\cite{Fayet:2006sp,McKeen:2009rm,Kamenik:2011vy,Darme:2020ral,Costa:2022pxv,Liu:2025lbw,Mescia:2026xju}. There are also recently some studies on the dark particle pair production from the flavor-conserving meson decay $\eta \to \pi^0 \chi \bar\chi$~\cite{Batell:2018fqo,Alvey:2019zaa,Flambaum:2020xxo,Su:2022wpj,PandaX:2023tfq,BESIII:2026qsu}.  



    \item $\nu^\ast \to \chi + {f} + \bar{f}$. Here $\nu^\ast$ is an off-shell neutrino, e.g. from SM particle decays. The typical processes of this category are the four-body meson decay ${\sf M}^- \to \ell^- \bar\chi \ell^{\prime+} \ell^{\prime-}$, four-body tauon decay $\tau^- \to \pi^- \chi \ell^+ \ell^-$ and five-body charged lepton decay $\ell^-_\gamma \to \ell_\alpha^- \ell_\beta^+ \ell_\beta^- \nu \bar\chi$. At the 1-loop order, the operator in \geqn{eqn:operator} could induce the exotic tauon decay $\tau^- \to  \pi^- \chi$~\cite{Davidson:2003ha}.
    
    
    \item $\chi + f \to \nu + f$. 
    In DM direct detection experiments and low threshold neutrino experiments, the effective operator in \geqn{eqn:operator} could induce the absorption of DM and convert all its mass into the kinematic energy~\cite{Dror:2019onn,Dror:2019dib,Hurtado:2020vlj,Chen:2021uuw,Ma:2024tkt,Ma:2024gqj}. Such a mechanism could significantly enhance the recoil energy in DM direct detection experiments, and as a result suitable for the searches of relatively light DM particles. High energy neutrinos can be produced from the scattering of relatively heavy DM with SM particles in the Sun or the Earth~\cite{Hurtado:2020vlj}.
    
    \item $\nu^{(\ast)} + f^{(\ast)} \to \chi + f$. 
    The dark fermion $\chi$ can be produced in neutrino scattering experiments.\footnote{A recent comprehensive EFT analysis of the coherent elastic neutrino-nucleus scattering (CE$\nu$NS) can be found in Ref.~\cite{Li:2026mco}.} The signal is the recoil energy of the target nuclei or electrons, as in DM direct detection experiments~\cite{Hurtado:2020vlj,Chen:2021uuw,Candela:2023rvt}. Such a process can also occur in the supernova core, which then leads to extra energy loss mechanisms and is thus constrained by the observations of SN1987A~\cite{Lin:2025mez}. The operator could also induce the exotic radiative decays of charged mesons ${\sf M}^- \to \ell^- \bar\chi$ and $W$ boson $W^- \to \ell^- \bar\chi$~\cite{Davidson:2003ha}.


    \item $\nu + f \to \nu + f$. Given the operator in \geqn{eqn:operator}, the dark fermion $\chi$ could contribute at the 1-loop level to the scattering of neutrinos with electrons or nuclei~\cite{Chao:2021bvq}. Such contributions will potentially interfere with the SM processes and distort the corresponding spectra. 

    \item $\nu +f \to \chi +f + f' + \bar{f}'$ and $\chi + f \to \nu + f + f' + \bar{f'}$, with $f'$ being a SM fermion. These trident processes can be induced by the operator in \geqn{eqn:operator}, which behave much like the neutrino trident process $\nu +f \to \nu +f + f' + \bar{f}'$~\cite{Altmannshofer:2014pba} and the dark trident process $\chi + f \to \chi + f + f' + \bar{f'}$~\cite{deGouvea:2018cfv,Dutta:2024nhg}.
\end{itemize}

In this paper, we take a general case for the fermion $\chi$ in the dimension-6 operator (\ref{eqn:operator}) to be a dark particle, e.g. some non-DM components in the dark sector. The fermion $\chi$ could also be a heavy sterile neutrino, e.g. from the type-I seesaw~\cite{Minkowski:1977sc,Mohapatra:1979ia,Yanagida:1979as,Gell-Mann:1979vob,Glashow:1979nm}. Generally speaking, the signals of the dark particle $\chi$ are quite similar to the DM case. However, there are still some differences: (i) The lifetime constraints on the dark particle $\chi$ is relaxed a lot, with respect to the DM case~\cite{Dror:2020czw,Ge:2022ius}. 
(ii) The DM absorption process $\chi + f \to \nu + f$ at DM direct detection experiments applies only to the searches of stable DM particles (at the cosmological time scale), but not to the dark particle $\chi$ that needs not to be stable.

There have been some studies of the effective operators in \geqn{eqn:operator} in the literature~\cite{Chang:2020jwl,Chen:2021uuw} and some investigations in the framework of simplified models with explicit mediators~\cite{Brdar:2018qqj,Chao:2021bvq,Candela:2023rvt,Candela:2024ljb}. 
If $\chi$ is identified as a heavy sterile neutrino, it can be produced from the heavy-light neutrino mixing~\cite{Kosmas:2017zbh,Blanco:2019vyp,Berryman:2019nvr,Miranda:2020syh,Alonso-Gonzalez:2023tgm,Behera:2023llq} (see Refs.~\cite{Bolton:2019pcu,sterileneutrino} for comprehensive lists of limits on heavy-light neutrino mixings) or from the neutrino magnetic moments~\cite{Vogel:1989iv,Balantekin:2013sda,Magill:2018jla,Shoemaker:2018vii,Brdar:2020quo,Schwetz:2020xra,Shoemaker:2020kji,Atkinson:2021rnp,Miranda:2021kre,Bolton:2021pey,DeRomeri:2022twg,Ovchynnikov:2022rqj}.
In this paper, we perform a comprehensive and systematic analysis of the dark particle $\chi$ production from the neutrinos scattering with nuclei targets in the experiments COHERENT and CONUS+ as well as with electrons in the DUNE near detector (ND), originating from the operators in \geqn{eqn:operator} with all the possible Lorentz structures. 
The corresponding limits on the corresponding cutoff scale $\Lambda$ can be mapped onto the simplified models with mediators (as long as the mediator mass is significantly larger than the involving momentum transferred), or even allow for a comparison with the ultraviolet completions in the literature~\cite{Dutta:2020scq}.

The main novelties and results of this paper are summarized as the following:
\begin{itemize}
    \item The latest COHERENT CsI and CONUS+ data are used to set limits on the hadronic absorption operators for all the possible Lorentz structures. In comparison, the COHERENT limits on the absorption operators in Ref.~\cite{Chang:2020jwl} are based on the early COHERENT data.
    \item The prospects of the leptonic absorption operators at the DUNE ND are estimated for all the possible Lorentz structures, which is absent in the literature to the best of our knowledge.
    \item All the existing limits on the hadronic absorption operators are collected in \gfig{fig:CEvNS sensitivity} and \gtab{tab:cevns}, i.e. those from the CE$\nu$NS experiments COHERENT and CONUS+ obtained in this paper, the Large Hadron Collider (LHC) data~\cite{Ma:2024tkt}, the $K$ and $B$ meson decays~\cite{Liu:2025lbw}, and the SN1987A observations~\cite{Lin:2025mez}. 
    It turns out that the constraints of the current COHERENT and CONUS+ data on the cutoff scales for the effective couplings could only go up to $\sim700$\,GeV at the 90\% confidence level (C.L.). The most stringent limits from the LHC data and SN1987A constraints can reach up to $\sim60$ TeV depending on the Lorentz structure~\cite{Ma:2024tkt,Lin:2025mez}.
    \item The existing limits on the leptonic absorption operators from CHARM II~\cite{Chen:2021uuw} and LEP~\cite{Fox:2011fx} are presented in \gfig{fig:DUNE sensitivity} and \gtab{tab:DUNE}. The cutoff scales for the couplings involving electrons could be probed up to $\sim 1$\,TeV at the 90\% C.L. by DUNE ND, beyond the existing CHARM II and LEP limits~\cite{Chen:2021uuw,Fox:2011fx}.  The sensitivities can be further improved up to tens of TeV at the future high-energy lepton colliders~\cite{Ge:2023wye}, such as the Future Circular Collider (FCC-ee)~\cite{FCC:2018evy}, Circular Electron-Positron Collider (CEPC)~\cite{CEPCStudyGroup:2018ghi}, International Linear Collider (ILC)~\cite{ILC:2013jhg,Bambade:2019fyw} and Compact Linear Collider (CLIC)~\cite{Aicheler:2018arh}. 
\end{itemize}

The rest of this paper is organized as follows. The EFT setup is sketched in \gsec{sec:EFT}, including some details on the decay of $\chi$ and the basic kinematic analysis. The full analysis details for the constraints of COHERENT and CONUS+ and the sensitivities of DUNE ND are presented in \gsec{sec:analysis}.
The probe of hadronic absorption operators at COHERENT
and CONUS+ are detailed in \gsec{sec:results:cevns}
while the DUNE ND sensitivities on the leptonic
operators are obtained in \gsec{sec:result:DUNE}.
In these two sections, we also compare our results to the main existing constraints and future prospects, and collect the relevant weaker constraints. We conclude in \gsec{sec:conclusion}. For the convenience of readers, the weak nuclear form factors and spin structure functions for the calculations in \gsec{subsec:benchmark} are collected in \gapp{app:nuclear}; the nucleon form factors are listed in \gapp{app:nucleon}, which are relevant to the content in \gsec{sec:LHC:cevns}.

\section{Neutrino scattering into a dark particle}
\label{sec:EFT}

\subsection{EFT setup}
\label{subsec:benchmark}

In the framework of EFT, the coupling of neutrinos $\nu_\alpha$ of flavor $\alpha = e,\, \mu,\, \tau$ with the dark fermion $\chi$ and the target fermion $f$ (electron $e$ or nucleon ${N}$) can be written as
\begin{align}
    \mathcal{L}_{\mathrm{eff}}=\sum_{i,f} \frac{{\cal O}_i}{\Lambda_{i,f}^2} ~+~ \mathrm{h.c.} \,,
    \label{lagrangian}
\end{align}
where $f = e,\, {N}$, 
and the index $i$ runs over all the five possible Lorentz structures for the new interactions: scalar ($S$), pseudo-scalar ($P$), vector ($V$), axial-vector ($A$) or tensor ($T$). These Lorentz structures correspond to the matrices of $\Gamma^i \equiv \{I,i\gamma^5,\gamma^\mu,\gamma^\mu\gamma^5,\sigma^{\mu\nu}\}$, respectively, with $\sigma^{\mu\nu}\equiv\frac{i}{2}[\gamma^\mu,\gamma^\nu]$. The parameter $\Lambda_{i, f}$ is an effective cutoff scale that characterizes the strength of the interaction: a smaller $\Lambda_{i, f}$ implies a stronger coupling. For simplicity, we have assumed that the effective couplings are flavor universal and conserving for all neutrinos $\nu_\alpha$. For the effective couplings involving electrons, the operators $(\bar\chi \Gamma^i P_L \nu) (\bar{e} \Gamma_i e)$ are correlated with the operator $(\bar{\nu} \Gamma^i e) (\bar{e} \Gamma_i \chi)$ via the Fierz transformations. The latter operator can originate from the dimension-6 operators involving heavy right-handed neutrinos, if we identify $\chi$ as a heavy neutrino~\cite{Beltran:2021hpq,Mitra:2022nri}. 

The interactions of $\nu_\alpha$ and $\chi$ with nucleons $N \equiv p,\,n$ can be constructed from the more fundamental interactions with the SM quarks, i.e. from the operators in the form of $(\bar{q}\Gamma^i q) (\chi \Gamma_i P_L \nu_\alpha)$~\cite{Shifman:1978zn,Drees:1993bu,Crivellin:2013ipa,Bertuzzo:2017lwt,Bishara:2017pfq}. 
The SM neutrinos $\nu_\alpha$ and dark particle $\chi$ could also couple to gluons at the 1-loop or higher order, which contributes also to the effective couplings of nucleons in \geqn{lagrangian}~\cite{Drees:1993bu,Hisano:2010fy,Hisano:2010ct,Hisano:2011cs,Hill:2011be,Hill:2014yka,Ibarra:2015fqa,Abe:2015rja,Abe:2018emu,Ertas:2019dew}. 
The energy scales for the effective couplings in \geqn{lagrangian} are characterized by the center-of-mass energies in the experiments COHERENT~\cite{COHERENT:2015mry}, CONUS+~\cite{CONUS:2024lnu} and DUNE~\cite{DUNE:2015lol},  
which are significantly below the electroweak scale of ${\cal O}(100\,{\rm GeV})$. At such low energy scales, there might also exist the couplings of $\chi$ with charged leptons $\ell \equiv e,\, \mu,\, \tau$, i.e. in the form of $\sim (\bar\chi \Gamma^i \ell) (\bar{f} \Gamma_i f)$, which, however, can be very different from those involving neutrinos in \geqn{lagrangian}. Focusing only on the couplings in \geqn{lagrangian}, we will not consider the corresponding couplings with charged leptons in this paper.




It is possible that the nucleon couplings depend on the isospin; in other words, the cutoff scales $\Lambda_{i,p}$ and $\Lambda_{i,n}$ could have different values. In some scenarios, the couplings of protons and neutrons may even have opposite signs~\cite{Feng:2011vu}. As a case study, we consider two distinct scenarios for the vector interaction, i.e. the isospin conserving (IC) one $\Lambda_{V, p} = \Lambda_{V,n}$ and the isospin violating (IV) one $\Lambda_{V, p} = -\Lambda_{V, n}$. For the other four Lorentz structure, i.e. the scalar, pseudo-scalar, axial-vector, and tensor interactions, we assume they are all of the IC type. 

\begin{figure}
\centering
\includegraphics[width=0.5\linewidth]{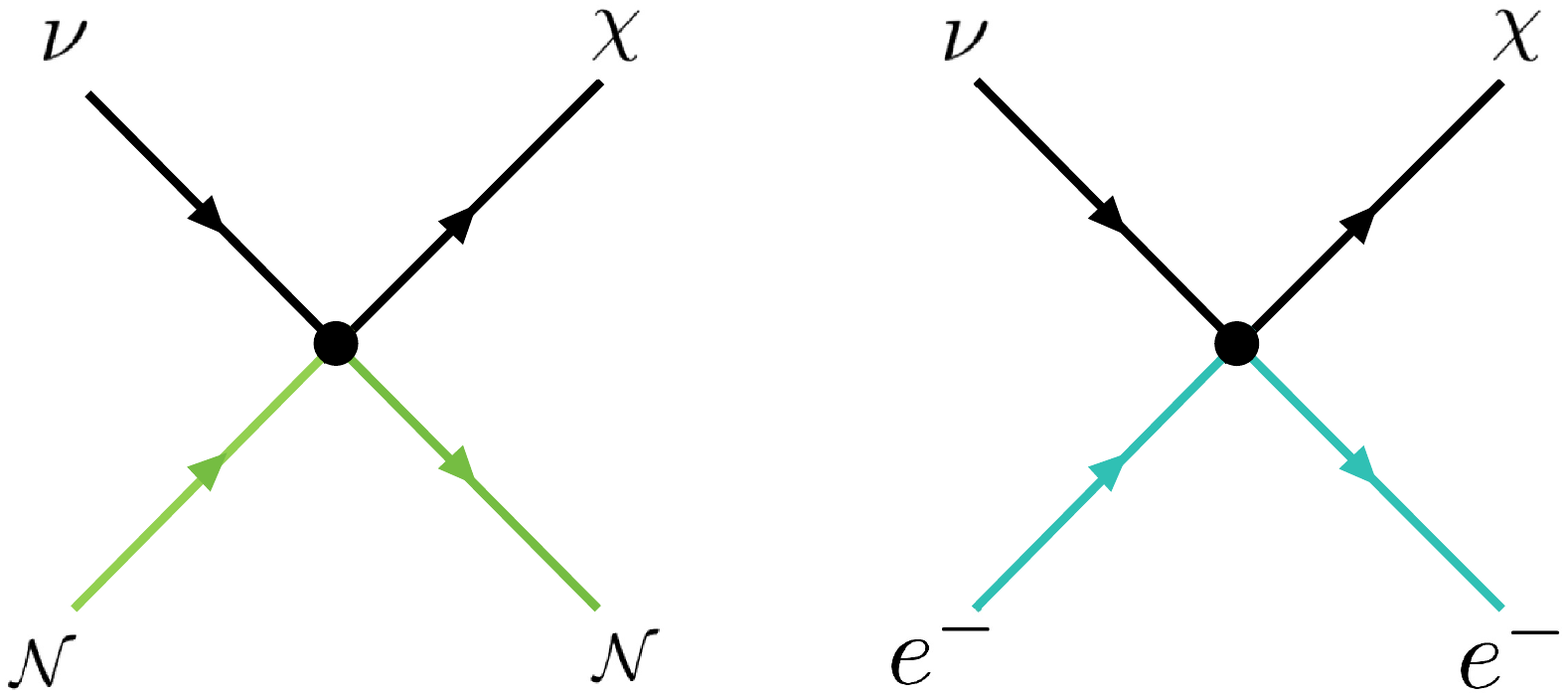}
\caption{Feynman diagrams for the neutrino scattering off nucleus $\nu_\alpha + {\cal N} \to \chi + {\cal N}$ (left) or electron $\nu_\alpha + e \to \chi + e$ (right), induced by the effective couplings in Eq.\,(\ref{lagrangian}).}
\label{fig:diagram}
\end{figure}

The effective couplings in \geqn{lagrangian} could induce the scattering of (anti)neutrinos $\nu_\alpha (\bar\nu_\alpha)$ with electrons $e^-$ or nuclei ${\cal N}$ in the target to produce the (anti) dark particle $\chi (\bar\chi)$, i.e. 
\begin{subequations}
\begin{align}
    \label{eqn:process:nue}
    \nu_{\alpha}(\bar\nu_\alpha) + e^{-} &\to \chi(\bar\chi) + e^{-} \,, \\
    \label{eqn:process:nuN}
    \nu_{\alpha}(\bar\nu_\alpha) + \mathcal{N} &\to \chi(\bar\chi) + \mathcal{N} \,.
\end{align}
\end{subequations}
The corresponding representative Feynman diagrams are shown in \gfig{fig:diagram}. While the effective couplings are assumed to be flavor universal, the experiments considered in this work are sensitive only to the (anti)neutrino flavors of $\nu_e$, $\bar{\nu}_e$, $\nu_\mu$, and $\bar{\nu}_\mu$.

For the scattering of neutrinos with nuclei, the calculations are a bit involved, as nuclei are composite particles. 
The nuclear structures have to be taken into account in the corresponding differential cross sections, which are interaction dependent. Within the nucleon level EFT description, the differential cross sections with respect to the nuclear recoil energy $T_{\cal N}$ can be written as~\cite{Freedman:1973yd,DeRomeri:2022twg}:
\begin{subequations}
\label{eqn:cs:N}
\begin{align}
\label{eqn:cs:N:S}
\frac{\mathrm{d}\sigma^{S}_{\mathcal{N}}}{\mathrm{d}T_\mathcal{N}} 
    &= \frac{{\sf A}^2 m_\mathcal{N}}{4\pi \Lambda_{S,{\cal N}}^4}F_W^2(|\bm{q}|^2)\left(1+\frac{T_\mathcal{N}}{2m_\mathcal{N}}\right)\left(\frac{m_\mathcal{N}T_\mathcal{N}}{E_\nu^2}+\frac{m_\chi^2}{2E_\nu^2}\right), \\
\label{eqn:cs:N:P}
\frac{\mathrm{d}\sigma^{P}_{\mathcal{N}}}{\mathrm{d}T_\mathcal{N}} 
    &= \frac{{\sf A}^2 m_{\mathcal{N}}}{4\pi \Lambda_{P,{\cal N}}^{4}}F_{W}^{2}(|\bm{q}|^{2})\frac{T_{\mathcal{N}}}{2m_{\mathcal{N}}}\left(\frac{m_{\mathcal{N}}T_{\mathcal{N}}}{E_{\nu}^{2}}+\frac{m_{\chi}^{2}}{2E_{\nu}^{2}}\right), \\
    \label{eqn:xs:N:V}
\frac{\mathrm{d}\sigma^{V}_{\mathcal{N}}}{\mathrm{d}T_\mathcal{N}} 
    &= \frac{C_V^2 m_{\mathcal{N}}}{2\pi \Lambda_{V,{\cal N}}^{4}}F_{W}^{2}(|\bm{q}|^{2})
    \left[\left(1-\frac{m_{\mathcal{N}}T_{\mathcal{N}}}{2E_{\nu}^{2}}-\frac{T_{\mathcal{N}}}{E_{\nu}}+\frac{T_{\mathcal{N}}^{2}}{2E_{\nu}^{2}}\right)-\frac{m_{\chi}^{2}}{4E_{\nu}^{2}}\left(1+\frac{2E_{\nu}}{m_{\mathcal{N}}}-\frac{T_{\mathcal{N}}}{m_{\mathcal{N}}}\right)\right], \\
    \label{eqn:cs:N:A}
\frac{\mathrm{d}\sigma^{A}_{\mathcal{N}}}{\mathrm{d}T_\mathcal{N}} 
    &= \frac{2m_\mathcal{N}}{(2J+1) \Lambda_{A,{\cal N}}^4} \left\{\left[\left(2+\frac{m_{\mathcal{N}}T_{\mathcal{N}}}{E_{\nu}^{2}}-\frac{2 T_{\mathcal{N}}}{E_{\nu}}-\frac{T_{\mathcal{N}}}{m_{\mathcal{N}}}+\frac{T_{\mathcal{N}}^{2}}{2 E_{\nu}^{2}}+\frac{T_{\mathcal{N}}^{2}}{m_{\mathcal{N}}E_{\nu}}\right)\right.\right. \notag \\
    &\quad -\left.\left.\frac{m_{\chi}^{2}}{2 E_{\nu}^{2}}\left(1+\frac{3E_{\nu}}{m_{\mathcal{N}}}+\frac{m_{\chi}^{2}}{m_{\mathcal{N}}T_{\mathcal{N}}}-\frac{T_{\mathcal{N}}}{2 m_{\mathcal{N}}}\right)\right]\tilde{S}^{\mathcal{T}}(|\bm{q}|^{2})\right. \notag \\
    &\quad + 2\left[\frac{T_{\mathcal{N}}}{E_{\nu}}\left(\frac{E_{\nu}}{m_{\mathcal{N}}}-\frac{T_{\mathcal{N}}}{2E_{\nu}}-\frac{T_{\mathcal{N}}}{m_{\mathcal{N}}}\right) + \left.\frac{m_{\chi}^{2}}{2E_{\nu}^{2}}\left(2+\frac{3E_{\nu}}{m_{\mathcal{N}}}+\frac{m_{\chi}^{2}}{m_{\mathcal{N}}T_{\mathcal{N}}}-\frac{T_{\mathcal{N}}}{2m_{\mathcal{N}}}\right)\right]\tilde{S}^{\mathcal{L}}(|\bm{q}|^{2})\right\}, \\
    \label{eqn:cs:N:T}
\frac{\mathrm{d}\sigma^{T}_{\mathcal{N}}}{\mathrm{d}T_\mathcal{N}} 
    &= \frac{m_\mathcal{N}}{(2J+1) \Lambda _{T,{\cal N}}^4} \left\{\left[\left(2-\frac{m_{\mathcal{N}}T_{\mathcal{N}}}{E_{\nu}^{2}}-\frac{2T_{\mathcal{N}}}{E_{\nu}}+\frac{T_{\mathcal{N}}}{m_{\mathcal{N}}}-\frac{T_{\mathcal{N}}^{2}}{m_{\mathcal{N}}E_{\nu}}-\frac{T_{\mathcal{N}}^{2}}{E_{\nu}^{2}}\right)\right.\right. \notag \\
    &\quad +\left.\left.\frac{m_{\chi}^{2}}{2E_{\nu}^{2}}\left(1+\frac{3E_{\nu}}{m_{\mathcal{N}}}-\frac{T_{\mathcal{N}}}{m_{\mathcal{N}}}+\frac{m_{\chi}^{2}}{m_{\mathcal{N}}T_{\mathcal{N}}}\right)\right]\tilde{S}^{\mathcal{T}}(|\bm{q}|^{2}) \right. \notag \\
    &\quad + \left[\left(1-\frac{T_{\mathcal{N}}}{E_{\nu}}-\frac{T_{\mathcal{N}}}{2m_{\mathcal{N}}}+\frac{T_{\mathcal{N}}^{2}}{2E_{\nu}^{2}}+\frac{T_{\mathcal{N}}^{2}}{2m_{\mathcal{N}}E_{\nu}}\right)\right. \notag \\
    &\quad -\left.\left.\frac{m_{\chi}^{2}}{2E_{\nu}^{2}}\left(1+\frac{3E_{\nu}}{2m_{\mathcal{N}}}-\frac{T_{\mathcal{N}}}{2m_{\mathcal{N}}}+\frac{m_{\chi}^{2}}{2m_{\mathcal{N}}T_{\mathcal{N}}}\right)\right]\tilde{S}^{\mathcal{L}}(|\bm{q}|^{2})\right\} \,,
\end{align}
\label{eq:dSigmadT}
\end{subequations}
where $m_{\mathcal{N}}$ is the nucleus mass, 
$J$ is the nuclear spin, ${\sf A}$ is the mass number of nucleus, and $C_{V}$ is the coefficient for the vector coupling. The function $F_W(|\bm{q}|^2)$ is the weak nuclear form factor accounting for the finite-size effects while $\tilde{S}^{\mathcal{T}}(|\bm{q}|^2)$ and $\tilde{S}^{\mathcal{L}}(|\bm{q}|^2)$ denote the transverse and longitudinal nuclear spin structure functions, respectively, which encode the spin-dependent nuclear response after combining proton and neutron couplings at the nucleon level. More details of the form factors and spin structure functions are collected in \gapp{app:nuclear}. 

It is transparent from \geqn{eqn:cs:N:S} and \geqn{eqn:cs:N:P} that, for scalar and pseudoscalar interactions, the scattering amplitudes add coherently over nucleons inside the target nucleus, leading to an overall enhancement proportional to ${\sf A}^2$, in close analogy with the CE$\nu$NS (Coherent Elastic Neutrino-Nucleus Scattering) processes. The corresponding (differential) cross sections are therefore enhanced by the atomic number ${\sf A}^2$ of target nuclei. 
At the nucleon level, the coherent effective charge $C_V$ depends on the relative sign between the proton and neutron couplings. So, for the vectorial couplings, the coefficient in \geqn{eqn:xs:N:V} is 
\begin{equation}
C_V = \begin{cases}
{\sf A} \,, & \text{isospin conserving} \,, \\
{\sf Z} - {\sf N} \,, & \text{isospin violating} \,,
\end{cases}
\end{equation}
with ${\sf Z}$ and ${\sf N}$ being the proton and neutron numbers inside the nucleus, respectively.
For the IV case with $\Lambda_{V,p} = - \Lambda_{V,n}$, the proton and neutron couplings cancel partially with each other in the coherent sum, 
which leads to a significant suppression of the cross section with respect to the IC case. As a result, the corresponding sensitivities of the IV cases depend critically on the neutron excess (${\sf N}-{\sf Z}$) of the target material. 
For the axial-vector and tensor interactions, as shown in \geqn{eqn:cs:N:A} and \geqn{eqn:cs:N:T},
the scattering is governed by the spin structures of the nucleus rather than by the total nucleon number. 

For the electron target, it is straightforward to calculate the differential cross sections, with respect to the recoil energy $T_e$ of electron~\cite{Chen:2021uuw,Candela:2024ljb}:
\begin{subequations}
\begin{align}
    \frac{{\rm d}\sigma_e^{S}}{{\rm d}T_{e}}
    &= \frac{m_{e}}{4\pi\Lambda_{S,e}^{4}}\left(1+\frac{T_{e}}{2m_{e}}\right)\left(\frac{m_{e}T_{e}}{E_{\nu}^{2}}+\frac{m_{\chi}^{2}}{2E_{\nu}^{2}}\right), \\
    \frac{{\rm d}\sigma_e^{P}}{{\rm d}T_{e}}
    &= \frac{m_{e}}{4\pi\Lambda_{P,e}^{4}}\frac{T_{e}}{2m_{e}}\left(\frac{m_{e}T_{e}}{E_{\nu}^{2}}+\frac{m_{\chi}^{2}}{2E_{\nu}^{2}}\right), \\
    \frac{{\rm d}\sigma_e^{V}}{{\rm d}T_{e}}
    &= \frac{m_{e}}{2\pi\Lambda_{V,e}^{4}}\left[\left(1-\frac{m_{e}T_{e}}{2E_{\nu}^{2}}-\frac{T_{e}}{E_{\nu}}+\frac{T_{e}^{2}}{2E_{\nu}^{2}}\right)-\frac{m_{\chi}^{2}}{4E_{\nu}^{2}}\left(1+\frac{2E_{\nu}}{m_{e}}-\frac{T_{e}}{m_{e}}\right)\right], \\
    \frac{{\rm d}\sigma_e^{A}}{{\rm d}T_{e}}
    &= \frac{m_{e}}{2\pi\Lambda_{A,e}^{4}}\left[\left(1+\frac{m_{e}T_{e}}{2E_{\nu}^{2}}-\frac{T_{e}}{E_{\nu}}+\frac{T_{e}^{2}}{2E_{\nu}^{2}}\right)+\frac{m_{\chi}^{2}}{4E_{\nu}^{2}}\left(1-\frac{2E_{\nu}}{m_{e}}+\frac{T_{e}}{m_{e}}\right)\right], \\
    \frac{{\rm d}\sigma_e^{T}}{{\rm d}T_{e}}
    &= \frac{4m_{e}}{\pi\Lambda_{T,e}^{4}}\left[\left(1-\frac{m_{e}T_{e}}{4E_{\nu}^{2}}-\frac{T_{e}}{E_{\nu}}+\frac{T_{e}^{2}}{4E_{\nu}^{2}}\right)-\frac{m_{\chi}^{2}}{4E_{\nu}^{2}}\left(\frac{1}{2}+\frac{2E_{\nu}}{m_{e}}-\frac{T_{e}}{2m_{e}}\right)\right].
\end{align}
\end{subequations}
where $m_e$ is the electron mass, $E_{\nu}$ is the incident neutrino energy, and $m_{\chi}$ is the dark particle mass. 
\subsection{Dark particle decay}

It is apparent in \geqn{lagrangian} that the stability of dark particle $\chi$ is not guaranteed, which, on contrary, leads inevitably to the decay of $\chi$. 
For the mass range of $m_\chi > 2m_e$, the interactions with electrons 
lead to the decay of $\chi$ into a neutrino and an electron-positron pair, i.e.
\begin{align}
    \chi \to \nu_\alpha+e^+ +e^- \,.
\end{align}
Among all the Lorentz structures, the tensor coupling gives the shortest 
lifetime and hence provides the most conservative estimate of the decay 
length. The corresponding partial width is~\cite{Ge:2023wye}
\begin{align}
    \label{eqn:width}
    \Gamma^T (\chi \to \nu_\alpha e^+e^-) =& \frac{m_\chi^5}{1024\pi^3\Lambda_{T,e}^4}
    \Big[ (16-56\eta-2\eta^2-3\eta^3)\sqrt{1-\eta} \nonumber\\
    & + 3 \eta^2 (16-\eta^2) {\rm arctanh} \sqrt{1-\eta} \Big] \,,
\end{align}
where $\eta \equiv 4m_e^2/m_\chi^2$. In the limit $m_\chi \gg 2m_e$, 
this reduces to $\Gamma^T \approx m_\chi^5/64\pi^3\Lambda_{T,e}^4$, 
which is heavily suppressed by the cutoff scale $\Lambda_{T,e}$. The 
corresponding proper decay length is
\begin{equation}
\tau^T_\chi \simeq 1.6\times 10^{9} \, {\rm cm} 
\left( \frac{m_\chi}{30\,{\rm MeV}} \right)^{-5}
\left( \frac{\Lambda_{T,e}}{1\,{\rm TeV}} \right)^{4} \,,
\end{equation}
which is much longer than the laboratory distances that we are interested in throughout this paper. The lifetimes for the other Lorentz structures are even longer since the tensor coupling gives the fastest decay rate among all the five operators~\cite{Ge:2023wye}. This indicates that the dark particle $\chi$ leaves the detectors without producing any visible signal for the entire parameter space in this work.

At the one-loop order, the visible decays of $\chi$ into a neutrino $\nu_\alpha$ plus photon(s) and the invisible decay of $\chi$ into three neutrinos are possible~\cite{Dror:2020czw,Ge:2022ius}:
\begin{equation}
\label{eqn:decays}
\chi \to \nu_\alpha \gamma \,, \quad
\chi \to \nu_\alpha \gamma\gamma \,, \quad
\chi \to \nu_\alpha \gamma\gamma\gamma \,, \quad
\chi \to \nu_\alpha \nu \bar\nu,
\end{equation}
depending on the Lorentz structure. 
These decays are also kinematically allowed for a light $\chi$ with mass below the MeV scale. The two-body decay $\chi \to \nu_\alpha \gamma$ is only allowed for the tensor coupling, while it is forbidden for other couplings by the charge conjugation symmetry of  quantum electrodynamics. In the tensor coupling case, the partial width is~\cite{Ge:2022ius}
\begin{equation}
\label{eqn:width:loop}
\Gamma^{T} (\chi \to \nu_\alpha \gamma) = \frac{\alpha m_e^2 m_\chi^3}{16\pi^4 \Lambda_{T,e}^4} \log^2 \left( \frac{\Lambda_{T,e}^2}{m_e^2} \right) \,,
\end{equation}
with $\alpha$ being the fine structure constant. The corresponding lifetime is also highly suppressed by the cutoff scale $\Lambda_{T,e}$ as well as the loop factor,
\begin{equation}
\label{eqn:lifetime:loop}
\tau^{T,e}_\chi \simeq 6\times 10^{11} \, {\rm cm} 
\left( \frac{m_\chi}{30\,{\rm MeV}} \right)^{-3}
\left( \frac{\Lambda_{T,e}}{1\,{\rm TeV}} \right)^{4}
\left( \frac{\log^2(\Lambda_{T,e}^2/m_e^2)}{1000} \right) \,.
\end{equation}
As detailed in Ref.~\cite{Ge:2022ius}, the lifetimes due to the three- and four-body decays at the 1-loop order in \geqn{eqn:decays} are much longer than the two-body decay $\chi \to \nu_\alpha \gamma$. 

Let us now move to the couplings of $\chi$ and neutrinos with nucleons. The dark particle $\chi$ in this paper has a mass below the GeV scale, and the decay channel $\chi \to \nu_\alpha NN$ is kinematically forbidden. All the 1-loop decays above can also be induced, with the electron in the loops replaced by a proton. 
Substituting the electron mass $m_e$ in \geqn{eqn:width:loop} by the nucleon mass $m_N$, the resultant lifetime is
\begin{equation}
\tau^{T,p}_\chi \simeq 6\times 10^{5} \, {\rm cm} 
\left( \frac{m_\chi}{30\,{\rm MeV}} \right)^{-3}
\left( \frac{\Lambda_{T,N}}{1\,{\rm TeV}} \right)^{4}
\left( \frac{\log^2(\Lambda_{T,N}^2/m_{N}^2)}{200} \right) \,,
\end{equation}
which is long enough for $\chi$ to decay outside the detectors for the parameter space of interest. 
Again, it is expected that with the couplings to nucleons, the partial widths of the three- and four-body decays in \geqn{eqn:decays} are suppressed by the phase space, with respect to the two-body decay. We note that for the nucleon tensor coupling, the loop-induced decay length $\tau^{T,p}_\chi$ could in principle be reduced to within the detector volume for $\Lambda_{T,N} \lesssim \mathcal{O}(200)$ GeV in our mass range. However, such low-$\Lambda_{T,N}$ regions are already excluded by the existing LHC data, which requires $\Lambda_{T,N} \gtrsim 1.1\,{\rm{TeV}}$ from the mono-$\gamma$ search (cf. ~\gfig{fig:CEvNS sensitivity} and ~\gtab{tab:cevns}). In short, for the purpose of this paper, the dark particle $\chi$ leaves the detectors without leaving any signal for the parameter space we are interested in, for both the couplings to electron and nucleons.






\subsection{Kinematic analysis}
\label{sec:kinematics}

The inelastic scattering process in \geqn{eqn:process:nue} and \geqn{eqn:process:nuN} features a kinematic threshold governed by the dark particle mass $m_\chi$ and the target mass $m_T$ ($m_e$ for electrons or $m_{\mathcal{N}}$ for nuclei).
For a given recoil kinetic energy $T_r$ of the target particle, the minimum incident neutrino energy required to produce $\chi$ is obtained from the four-momentum conservation and the mass on-shell conditions~\cite{Candela:2024ljb}:
\begin{align}
  E_\nu^{\min}(m_\chi, T_r)
=
  \frac 1 2
\left(
  T_r
+ \sqrt{2 m_T T_r + T^2_r}
\right)
\left(
  1
+ \frac {m_\chi^2}{2 m_T T_r}
\right).
\label{eq:Emin}
\end{align}
Here $\sqrt{2 m_T T_r + T^2_r} = |\bm p'|$ is the three-momentum of the recoiling target particle, which implies that the required energy depends on both the dark particle mass $m_\chi$ and the recoil energy $T_r$. 
This expression reduces to the well known elastic scattering limit when $m_\chi = 0$. 
Conversely, for a fixed neutrino energy $E_\nu^{}$, the kinematically allowed recoil energy $T_r$ is bounded. Solving the energy and momentum conservation equations for $T_r$ gives the maximum and minimum recoil energies:
\begin{subequations}
\begin{align}
  T_{\max} (m_\chi,\, E_\nu)
& \ = \
  \frac{2 m_T E_\nu^2 - m_\chi^2 (E_\nu^{} + m_T) + E_\nu^{} \sqrt{\Delta} }{2 m_T (2 E_\nu^{} + m_T)} \,,
\\
  T_{\min} (m_\chi,\, E_\nu)
& \ = \
  \frac{2 m_T E_\nu^2 - m_\chi^2 (E_\nu^{} + m_T) - E_\nu^{} \sqrt{\Delta} }{2 m_T (2E_\nu^{} + m_T)} \,,
\label{eq:Tminmax}
\end{align}
\end{subequations}
where 
\begin{equation}
  \Delta
\equiv
  4 m^2_T E_\nu^2
- 4 m_T m_\chi^2 (E_\nu^{} + m_T)
+ m_\chi^4 \,.
\end{equation}
These relations apply to the scatterings involving both electrons and nuclei.









The kinematic upper bound on the dark particle mass $m_\chi$ is determined by energy momentum conservation and the maximal available neutrino energy. In the limit of vanishing recoil energy $T_r \to 0$, the kinematic upper bound is~\cite{Candela:2024ljb} 
\begin{align}
  m_\chi
\lesssim
  \sqrt{m_T (m_T + 2 E_\nu)} - m_T \,,
\label{eq:mass_limit}
\end{align}
This bound corresponds to the point of $T_{\min}=T_{\max}$, and therefore represents the maximal dark particle mass that can be produced. 
This kinematic constraint allows to estimate the maximal dark particle  mass that can be probed at different neutrino experiments. At the DUNE experiment, the neutrino energy can reach ${\cal O} (10\,{\rm GeV})$~\cite{DUNE:2015lol}, which implies sensitivity to dark particle masses up to $\mathcal{O}(100)$\,MeV.
For COHERENT and CONUS+ experiments, the nuclear target masses are much larger than the neutrino energies. In this regime, \geqn{eq:mass_limit} can be simplified to $m_\chi \lesssim E_\nu^{\rm max}$, indicating that almost the entire neutrino energy carried by the incoming neutrino is nearly all converted to be the dark particle mass in the final state. For COHERENT at the SNS, the neutrino energies are about 50\,MeV~\cite{COHERENT:2015mry}. In CONUS+, reactor  antineutrinos exhibit lower energies below $\sim 12\,{\rm MeV}$~\cite{CONUS:2024lnu}. 




The typical momentum transfer in the scattering process, $|\bm q| \simeq \sqrt{2 m_T T_r}$ plays an important role in assessing the validity of the EFT description. In the framework of EFT the cutoff scale $\Lambda$ is required to satisfy $\Lambda \gg |\bm q|$ to ensure the point-like interaction approximation. It turns out that the experiments COHERENT, CONUS+ and DUNE ND probe distinct momentum transfer regimes due to their different targets and recoil energy sensitivities: 
\begin{itemize}
\item 
For the spallation source CE$\nu$NS experiment COHERENT, the target is CsI, with nuclear masses $m_{\rm Cs}\simeq 124\,{\rm GeV}$ and $m_{\rm I}\simeq 118\,{\rm GeV}$, and typical nuclear recoil energies $T_{\mathcal{N}}\sim$ few tens of keV. This implies a characteristic momentum transfer $|\bm q|\simeq \sqrt{2m_{\mathcal N}T_{\mathcal N}} \sim {\cal O}(10\!-\!100)\,{\rm MeV}$~\cite{COHERENT:2015mry}.

\item For the reactor neutrino experiment CONUS+,  with a germanium target $m_{\rm Ge} \simeq 66\,{\rm GeV}$ and nuclear recoil energies $T_{\mathcal{N}} \sim {\cal O}(0.1\!-\!1)\,{\rm keV}$, the characteristic momentum transfer is $\lvert\bm q \rvert \sim {\cal O}(1\!-\!10)\,{\rm MeV}$~\cite{CONUS:2024lnu}.

\item In the DUNE experiment, the accelerator neutrino energies are much higher, and the electron recoil energy $T_{e}$ ranges from $\sim 30\,\text{MeV}$ to several GeV~\cite{DUNE:2015lol}. 
However, for the momentum transfer is given by $|\bm q| \simeq \sqrt{2 m_e T_e}$. As a result, even for electron recoil energies at the GeV scale, the momentum transfer remains well below the GeV scale, typically $\mathcal{O}(10\!-\!50)\,\text{MeV}$.
\end{itemize}
As we will see in \gfig{fig:CEvNS sensitivity} and \gfig{fig:DUNE sensitivity}, the COHERENT and CONUS+ limits on the cutoff scales and the prospects at DUNE ND are all well above the corresponding characteristic transferred momenta, unless the dark particle mass is close to the kinematic threshold, where larger recoil energies (and hence larger $|\bm q|$) are required. This indicates that the EFT description in this paper is valid. This is further quantified in \gtab{tab:EFT validity}, which summarizes the maximal recoil energy $T_\text{max}$, the corresponding maximal momentum transfer $q_\text{max}$, the weakest limit or prospect of the cutoff scale $\Lambda_\text{min}$ at the 90\%\,C.L., and the maximal dark particle mass $m_{\chi,\,\rm max}$ for the experiments COHERENT, CONUS+ and DNUE ND. 
Here $m_{\chi,\,\rm max}$ is an estimate of the dark particle mass at which the EFT validity condition $\Lambda \gg |\bm q|$ begins to break down, i.e. $\Lambda_{\rm min} \simeq q_{\rm max}$. For the dark particle mass approaching $m_{\chi,\,\rm max}$, not only does the EFT description become unreliable, but the available phase space is also severely reduced, leading to a rapid weakening of the experimental sensitivities, as clearly seen in \gfig{fig:CEvNS sensitivity} and \gfig{fig:DUNE sensitivity}.

\begin{table}[!t]
\centering
\caption{The maximal recoil energy $T_{\rm max}$,  
the corresponding maximal momentum transfer $q_\text{max}$, the weakest limit or prospect of the cutoff scale $\Lambda_\text{min}$ at the 90\%\,C.L., and the maximal dark particle mass $m_{\chi,\,\rm max}$ for the experiments COHERENT, CONUS+ and DNUE ND. 
If the dark particle mass is at $m_{\chi,\,{\rm max}}$, we have $\Lambda_\text{min} \simeq q_\text{max}$, and the EFT validity condition $\Lambda \gg |\bm{q}|$ begins to break down. See text for more details.
}
\label{tab:EFT validity}
\vspace{5pt}
\begin{tabular}{c|c|c|c|c|c}
\toprule
Experiment & Target & $T_\text{max}$ & $q_\text{max}$ & $\Lambda_\text{min}$ & $m_{\chi,\,\rm max}$ \\
\midrule
COHERENT & CsI & $\sim 50\,\text{keV}$ & $\sim 110\,\text{MeV}$ & $8.2\,\text{GeV}$ & $ 55\,\text{MeV}$\\ \hline
CONUS+ & Ge & $\sim 1\,\text{keV}$ & $\sim 12\,\text{MeV}$ & $7.5\,\text{GeV}$ & $ 7.5\,\text{MeV}$ \\ \hline
DUNE ND & $e^-$ & $\sim 3\,\text{GeV}$ & $\sim 55\,\text{MeV}$ & $560\,\text{GeV}$ & $ 99\,\text{MeV}$ \\
\bottomrule
\end{tabular}
\end{table}





\section{Analysis details}
\label{sec:analysis}

In this section, we provide all the analysis details for estimating the limits on the cutoff scales from the current COHERENT data in \gsec{coherent} and the CONUS+ data in \gsec{subsec:conusplus}. The analysis details for the DUNE ND sensitivities are given in \gsec{sec:analysis:dune}.




\subsection{COHERENT}
\label{coherent}


We analyze the CsI detector dataset released by the COHERENT collaboration in 2021~\cite{COHERENT:2021xmm}.
The CsI detector has a fiducial mass $m_{\rm det}=14.57\,\mathrm{kg}$ and is located at a distance
$L=19.3\,\mathrm{m}$ from the Spallation Neutron Source (SNS)~\cite{COHERENT:2017ipa,COHERENT:2021xmm}.
Neutrinos are produced via pion decay at rest, giving rise to a pulsed flux consisting of a prompt
mono-energetic $\nu_\mu$ component from $\pi^+\to\mu^+\nu_\mu$, followed by delayed $\nu_e$ and
$\bar\nu_\mu$ components from muon decay.
The neutrino energy spectra and flavor dependent timing distributions are taken from the official COHERENT CsI-2021 data release~\cite{Louis:2009zza,COHERENT:2021xmm}. A similar study can be found in Ref.~\cite{Chang:2020jwl}, which is, however based on the early COHERENT data~\cite{COHERENT:2017ipa}. 

The experimental observables are the reconstructed number of photoelectrons $n_{\rm PE}$ and the
reconstructed arrival time $t_{\rm rec}$.
Following the COHERENT collaboration, we perform a binned two-dimensional analysis in the
$(n_{\rm PE},t_{\rm rec})$ plane~\cite{COHERENT:2021xmm,DeRomeri:2022twg,Candela:2023rvt}. For a given neutrino flavor $\nu_\alpha$ and nuclear target $\mathcal N=\mathrm{Cs}$ or $\mathrm{I}$,
the differential event rate with respect to the nuclear recoil energy $T_{\mathcal N}$ is
\begin{equation}
\frac{{\rm d}N_{\nu_\alpha\mathcal N}}{{\rm d}T_{\mathcal N}}
=
N_{\rm target}
\int_{E_\nu^{\rm min}(T_{\mathcal N})}^{E_\nu^{\rm max}} {\rm d} E_\nu\;
\frac{{\rm d} \Phi_{\nu_\alpha}}{{\rm d} E_\nu}(E_\nu)\,
\frac{{\rm d} \sigma_{\nu_\alpha\mathcal N}}{{\rm d} T_{\mathcal N}}(E_\nu,T_{\mathcal N}) \,,
\label{eq:coh_recoil_rate}
\end{equation}
where $N_{\rm target} \equiv N_A m_{\rm det}/M_{\mathcal N}$ is the number of target nuclei.
The contributions from the Cs and I nuclei are computed separately and combined
according to their corresponding target numbers in the CsI crystal.

The nuclear recoil energy is converted into an electron-equivalent energy $E_{ee}$ through the
energy dependent quenching factor $Q(T_{\cal N})$ via
\begin{equation}
E_{ee} = Q(T_{\mathcal N})\,T_{\mathcal N},
\end{equation}
where $Q(T_{\mathcal N})$ is taken from the COHERENT CsI-2021 release and parameterized as a
fourth-order polynomial using the neutron calibration data~\cite{COHERENT:2021xmm}.
The electron-equivalent energy is subsequently converted into the mean number of photoelectrons
using a fixed light yield,
\begin{equation}
  \bar n_{\rm PE} = Y\,E_{ee},
\quad \text{with} \quad
  Y = 13.35\,\mathrm{PE/keV_{ee}} \,.
\end{equation}
The detector effects, including photoelectron statistics, energy resolution, and the
$n_{\rm PE}$-dependent detection efficiency $\epsilon_E(n_{\rm PE})$, are implemented through
the response kernel $\mathcal R(n_{\rm PE}\mid T_{\mathcal N})$ following the official
COHERENT CsI-2021 prescription~\cite{COHERENT:2021xmm,DeRomeri:2022twg,Candela:2023rvt}.
The SNS neutrino flux exhibits a pronounced time structure. 
For each neutrino flavor $\nu_\alpha$, we adopt the arrival time probability
density $P_T^{\nu_\alpha}(t)$ by the COHERENT collaboration, and the reconstructed time $t_{\rm rec}$ is implemented together
with the time dependent efficiency $\epsilon_T(t_{\rm rec})$ following the CsI-2021 prescription. The explicit functional forms and numerical parameters entering
$\mathcal R(n_{\rm PE}\mid T_{\mathcal N})$, $\epsilon_E$, $P_T^{\nu_\alpha}$, and $\epsilon_T$
are taken from the supplemental material of Ref.~\cite{COHERENT:2021xmm}.

The expected number of CE$\nu$NS events in the $(i,j)$-th bin of $(n_{\rm PE},t_{\rm rec})$
is given by
\begin{align}
{\cal N}^{\rm CE\nu NS}_{ij}
\ = \ & \sum_{\nu_\alpha} \sum_{\mathcal N=\mathrm{Cs,I}}
\int {\rm d}T_{\mathcal N}\;
\frac{{\rm d}N_{\nu_\alpha\mathcal N}}{{\rm d}T_{\mathcal N}}
\int {\rm d}n_{\rm PE}\;
\mathcal I_i(n_{\rm PE})\,
\epsilon_E(n_{\rm PE})\,
\mathcal R(n_{\rm PE}\mid T_{\mathcal N})
\nonumber
\\
\times &
  \int {\rm d}t_{\rm rec}\;
  \mathcal I_j(t_{\rm rec})\,
  \epsilon_T(t_{\rm rec})\,
  P_T^{\nu_\alpha}(t_{\rm rec}) \,,
\label{eq:coh_Nij_CEvNS}
\end{align}
where $\mathcal I_i$ and $\mathcal I_j$ denote the indicator functions for the $i$-th
$n_{\rm PE}$ bin and the $j$-th time bin, respectively.
In the CsI-2021 analysis, the reconstructed photoelectron axis is divided into 9 bins and the
reconstructed time axis is divided into 11 bins, yielding a total of $9\times 11$ bins.
The bin boundaries follow exactly those adopted in the official COHERENT CsI-2021 release~\cite{COHERENT:2021xmm}.
All expected event numbers entering the likelihood analysis are evaluated in the corresponding $(n_{\rm PE},t_{\rm rec})$ bins. 

The experimental background components, including the beam-related neutrons (BRN),
neutrino-induced neutrons (NIN), and steady-state background (SSB), are taken directly
from the templates provided by the COHERENT collaboration~\cite{COHERENT:2021xmm}.
Although the elastic neutrino-electron scattering (E$\nu$ES) may in principle contribute
due to the lack of recoil discrimination in CsI, we do not model E$\nu$ES as an independent component, since its impact is subdominant for the CsI-2021 analysis window and does not affect too much the limits on the NP parameters.
This effect can be effectively absorbed into the background normalization and systematic
uncertainties~\cite{DeRomeri:2022twg,Candela:2023rvt}.

\begin{figure}[t]
\centering
\includegraphics[width=0.7\linewidth]{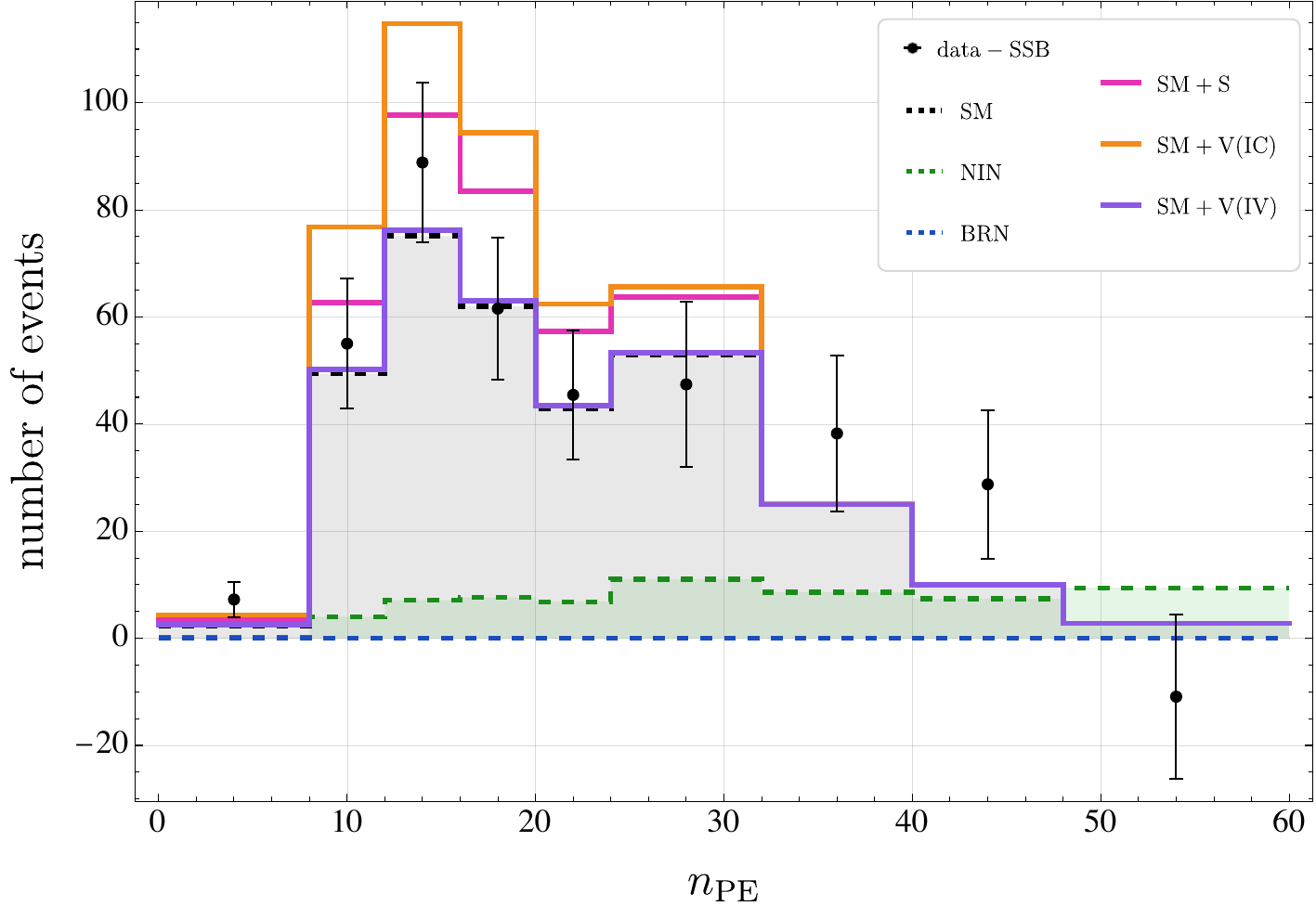}
\caption{Event numbers at the COHERENT experiment
(CsI-2021 dataset) for the SM CE$\nu$NS process
(black dashed), the NIN (green dashed) and BRN
(blue dashed) backgrounds as well as the NP
scenarios with the scalar (pink solid), IC
(orange solid) or IV (purple solid) vector coupling,
as functions of $n_{\rm PE}$. The data with error bars are for the residual events after subtracting the SSB background. The backgrounds are labeled by the dashed lines with shaded regions, while the NP cases are depicted as the solid lines. We have set the parameters $m_\chi = 10\,\mathrm{MeV}$ and $\Lambda = 500\,\mathrm{GeV}$.}
\label{fig:coherent_pe_projection}
\end{figure}

The event distributions for the SM CE$\nu$NS process and the BRN and NIN backgrounds for different values of $n_{\rm PE}$ are presented as the dashed lines with shaded regions in \gfig{fig:coherent_pe_projection}, where we have integrated over the reconstructed time.
For illustration purpose, the expected events with the NP scenarios of the scalar, IC vector and IV vector couplings are depicted as the solid purple, green and blue lines, respectively. We have taken $m_\chi = 10$\,MeV and $\Lambda_{i,N} = 500$\,GeV for all these scenarios. 
The data with errors are for the events after subtracting the SSB background.
It is clear in \gfig{fig:coherent_pe_projection} that the NP contributions could significantly affect the neutrino-nucleus scattering at the COHERENT experiment and are thus constrained by the experimental data, depending on the interaction type and cutoff scale.

Based on the full two-dimensional $(n_{\rm PE},t_{\rm rec})$ distributions described above,
the total theoretical prediction within each $(i,j)$ bin can be written as 
\begin{align}
{\cal N}^{\rm th}_{ij}
\ = \ & 
(1+\alpha_0+\alpha_5)\,({\cal N}^{\rm CE\nu NS}_{ij}+{\cal N}^{\rm NP}_{ij})
\nonumber
\\
+ \ &
  (1+\alpha_1)\,{\cal N}^{\rm BRN}_{ij}
+ (1+\alpha_2)\,{\cal N}^{\rm NIN}_{ij}
+ (1+\alpha_3)\,{\cal N}^{\rm SSB}_{ij} \,,
\label{eq:coh_Nth}
\end{align}
where ${\cal N}_{ij}^{\rm CE \nu NS}$ stands for the SM CE$\nu$NS contribution~\cite{Freedman:1973yd,Drukier:1984vhf}, and ${\cal N}_{ij}^{\rm NP}$ is for the NP contribution due to the absorption operators in \geqn{lagrangian}. The nuisance parameters $\alpha_k$ account for the dominant systematic uncertainties:
$\sigma_0=11\%$ for the overall flux and efficiency normalization,
$\sigma_1=25\%$, $\sigma_2=35\%$, and $\sigma_3=2.1\%$ for the BRN, NIN, and SSB normalizations, respectively, while
$\sigma_5=3.8\%$ is for the quenching factor uncertainty.
Additional nuisance parameters associated with the nuclear form factor and timing uncertainties
are treated following the COHERENT CsI-2021 prescription~\cite{COHERENT:2021xmm,DeRomeri:2022twg}. 


To set limits on the NP contributions, we construct a binned Poisson likelihood in the two-dimensional $(n_{\rm PE},t_{\rm rec})$ space~\cite{COHERENT:2021xmm,DeRomeri:2022twg}:
\begin{equation}
\chi^2=2\sum_{i,j}\left[{\cal N}^{\rm th}_{ij}-{\cal N}^{\rm obs}_{ij}+{\cal N}^{\rm obs}_{ij}\log\!\left(\frac{{\cal N}^{\rm obs}_{ij}}{{\cal N}^{\rm th}_{ij}}\right)\right]+\sum_k \left(\frac{\alpha_k}{\sigma_k}\right)^2 ,
\label{eq:coh_chi2}
\end{equation}
where ${\cal N}^{\rm obs}_{ij}$ denotes the observed number of events. 
All the nuisance parameters are profiled to derive constraints on the NP parameters.

\subsection{CONUS+}
\label{subsec:conusplus}

\begin{figure}
\centering
\includegraphics[width=0.75\linewidth]{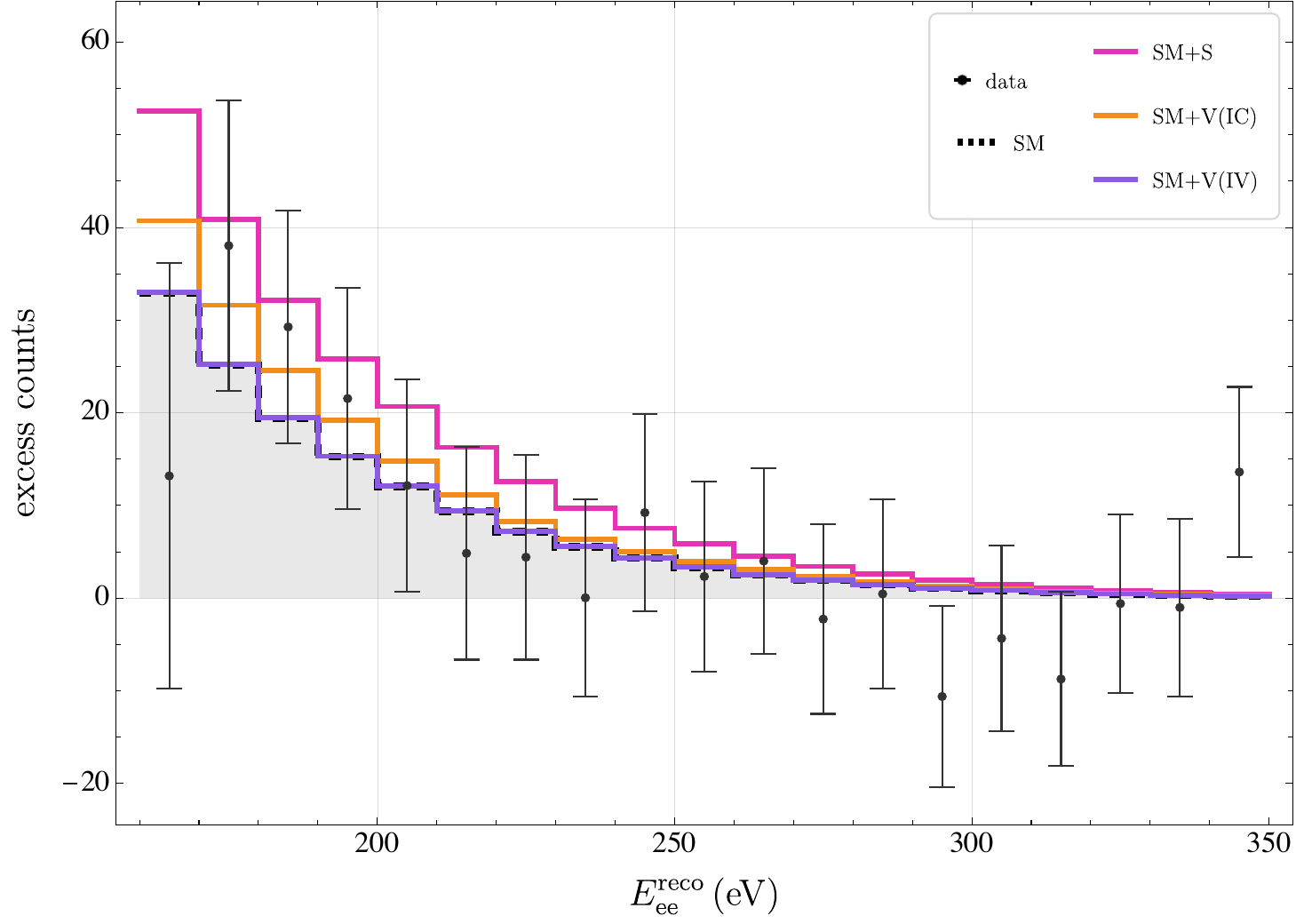}
\caption{Binned CONUS+ excess spectra as functions of the reconstructed  energy $E_{\rm ee}^{\rm reco}$, for the SM CE$\nu$NS process (dashed black line with shaded region), and the NP scenarios with the scalar, IC or IV vector coupling (solid colored lines), as functions of $E_{\rm ee}^{\rm reco}$. The data with error bars are the measured spectrum. We have set the parameters $m_\chi = 5\,\mathrm{MeV}$ and $\Lambda = 500\,\mathrm{GeV}$.}
\label{fig:conus_event_num}
\end{figure}

The CONUS+ experiment operates at a baseline of $L\simeq 20.7\,\mathrm{m}$ from the Leibstadt nuclear power plant in Switzerland, which provides a high reactor $\bar\nu_e$ flux at the detector location of $\phi_{\bar\nu_e}\simeq 1.5\times 10^{13}\,\mathrm{cm}^{-2}\,\mathrm{sec}^{-1}$.
The detector array consists of three high-purity germanium (HPGe) detectors (C2, C3, C5) with fiducial masses $0.95\,\mathrm{kg}$, $0.94\,\mathrm{kg}$ and $0.94\,\mathrm{kg}$, respectively, amounting to a total active mass of $2.83\pm 0.02\,\mathrm{kg}$. The corresponding analysis thresholds are $T_e^{\rm th}=160\,\mathrm{eV_{ee}}$ (C3), $170\,\mathrm{eV_{ee}}$ (C5) and $180\,\mathrm{eV_{ee}}$ (C2)~\cite{Ackermann:2025obx, Chattaraj:2025fvx}. After data quality selections, the exposure amounts to $327\,\mathrm{kg\cdot day}$ (reactor on) and $60\,\mathrm{kg\cdot day}$ (reactor off)~\cite{Ackermann:2025obx}. The low energy thresholds and high reactor neutrino flux make CONUS+ particularly sensitive to low-energy neutrino interactions, including the SM CE$\nu$NS process and possible NP contributions in the nuclear recoil channel.

In this work, we follow the one-dimensional spectral analysis strategy \cite{DeRomeri:2025csu} based on the reconstructed ionization energy spectrum. In the estimates of the CONUS+ limits of the NP contributions, 
we adopt a CE$\nu$NS-only treatment, which is consistent with the fact that the NP effects at the CONUS+ experiments are dominated by the nuclear channel. 
The CE$\nu$NS+E$\nu$ES choice can lead to small differences in the light mass regime in some NP cases~\cite{DeRomeri:2025csu}.

The predicted differential rate in the reconstructed ionization energy $E_e^{\rm reco}$ is obtained by convolving the reactor $\bar\nu_e$ flux with the differential cross section and the detector response, 
\begin{align}
  \frac{{\rm d}\mathcal{R}}{{\rm d}E_{\rm ee}^{\rm reco}} =\mathcal{E}\int {\rm d}T_{\mathcal N}\; {\cal G} (E_{\rm ee}^{\rm reco},E_{\rm er})\int {\rm d}E_\nu\; \frac{{\rm d}\phi}{{\rm d}E_\nu}\,\frac{{\rm d}\sigma_{\bar\nu_e \mathcal N}}{{\rm d}T_{\mathcal N}} \,,  
\end{align}
where $\mathcal{E}$ denotes the exposure, and ${\cal G}$ is the energy resolution kernel~\cite{DeRomeri:2025csu, Chattaraj:2025fvx}, \( E_{\mathrm{er}} \) is the true electron-equivalent recoil energy, and the integration bounds of the nuclear recoil energy are determined by the detector threshold through the quenching relation, and by the kinematic maximum recoil energy for a given neutrino energy. The energy resolution function is modeled as a Gaussian distribution,
\begin{equation}
  \mathcal{G}(E_{\mathrm{ee}}^{\mathrm{reco}}, E_{\mathrm{er}})
\equiv
  \frac{1}{\sqrt{2\pi}\sigma_{\mathrm{res}}}
  \exp
\left[
- \frac {(E_{\mathrm{ee}}^{\mathrm{reco}} - E_{\mathrm{er}})^2}
        {2\sigma_{\mathrm{res}}^2}
\right] \,,
\end{equation}
with the resolution width given by~\cite{Lindner:2024eng,Chattaraj:2025fvx}:
\begin{equation}
  \sigma_{\mathrm{res}}
\equiv
  \sqrt{\sigma_0^2 + \mathcal{F}_{\mathrm{fano}} \eta E_{\mathrm{er}}} \,.
\end{equation}
Here \( \sigma_0 = 20.38\,\mathrm{eV_{ee}} \) (derived from a full width at half maximum of $48\,\mathrm{eV_{ee}}$.), \( \mathcal{F}_{\mathrm{fano}} = 0.1096 \) is the Fano factor for germanium~\cite{Ackermann:2025obx}, and \( \eta = 2.96\,\mathrm{eV_{ee}} \) is the average energy required to create an electron-hole pair. 
The true nuclear recoil energy \( T_{\mathcal{N}} \) is related to the electron-equivalent energy \( E_{\mathrm{er}} \) via the quenching factor ${Q}_{\rm QF}$:
\begin{equation}
E_{\mathrm{er}} \ = \ Q_{\rm QF}(T_{\mathcal{N}}) T_{\mathcal{N}} \ = \ \frac{k g(\epsilon)}{1 + k g(\epsilon)} T_{\mathcal{N}} \,,
\end{equation}
where \( \epsilon = 11.5 {\sf Z}^{-7/3} T_{\mathcal{N}} \), \( g(\epsilon) = 3\epsilon^{0.15} + 0.7\epsilon^{0.6} + \epsilon \), and \( k = 0.162 \pm 0.004 \) is the Lindhard parameter~\cite{Bonhomme:2022lcz}. Using this relation, the differential cross section with respect to $E_{\rm cr}$ can be straightforwardly obtained from the nuclear recoil cross section in \geqn{eqn:cs:N}. 
The reactor antineutrino spectrum ${\rm d}\phi / {\rm d} E_\nu$ is constructed using the spectral function from Ref.~\cite{Kopeikin:2012zz} for \( E_\nu < 2\,\mathrm{MeV} \) and from Ref.~\cite{Mueller:2011nm} for higher energies. 


The predicted number of events in each reconstructed energy bin is:
\begin{equation}
{\cal R}_i^{\mathrm{th}} = \int_{i} \frac{\mathrm{d} {\cal R}}{\mathrm{d}E_{\mathrm{ee}}^{\mathrm{reco}}} \, \mathrm{d}E_{\mathrm{ee}}^{\mathrm{reco}} \,,
\end{equation}
which includes both the SM CE$\nu$NS and NP contributions, i.e.
\begin{align}
    {\cal R}_i^{\mathrm{th}} ={\cal R}_i^{\mathrm{CE\nu NS}} + {\cal R}_i^{\rm NP} \, .
\end{align}
For illustration purpose, the spectra for the pure SM process and those for the scalar, IC vector and IV vector NP cases are shown in \gfig{fig:conus_event_num} as the dotted and solid purple, green and blue lines, respectively, as functions of the reconstructed energy $E_{\rm ee}^{\rm reco}$. For the NP scenarios we have taken the benchmark values of $m_\chi = 5$\,MeV and $\Lambda_{i,N} = 500$\,GeV. As in \gfig{fig:coherent_pe_projection}, the spectra might be disturbed by the NP $\bar\nu_e\mathcal{N} \to \bar\chi \mathcal{N}$ scattering. 


The statistical analysis is performed by using the \( \chi^2 \) function~\cite{Lindner:2024eng,Chattaraj:2025fvx}:
\begin{equation}
\chi^2 = \sum_{i}^{} \frac{\left[ {\cal R}_i^{\mathrm{exp}} - (1 + \alpha) {\cal R}_i^{\mathrm{th}} \right]^2}{\sigma_i^2} + \left( \frac{\alpha}{\sigma_\alpha} \right)^2 \,,
\end{equation}
where \( \alpha \) is a nuisance parameter accounting for the combined systematic uncertainty with \( \sigma_\alpha = 16.9\% \). This uncertainty incorporates contributions from the reactor antineutrino flux (4.6\%), quenching factor (7.3\%), energy threshold (14.1\%), active detector mass (1.1\%), trigger efficiency (0.7\%), and nuclear form factor (3.2\%)~\cite{Lindner:2024eng,Chattaraj:2025fvx}. The CONUS+ collaboration reports \( 395 \pm 106 \) CE$\nu$NS events~\cite{Chattaraj:2025fvx}, consistent with the SM prediction of \( 347 \pm 59 \) events, which allows us to derive constraints on the NP parameter space. We note that the germanium quenching factor at sub-keV recoil energies is subject to significant uncertainties~\cite{Colaresi:2022obx,Kavner:2024xxd}, which can impact significantly the results on new physics~\cite{Li:2025pfw}. In our analysis, we adopt the Lindhard model with $k=0.162\pm0.004$~\cite{Bonhomme:2022lcz}, consistent with the quenching factor measurement reported by the CONUS+ collaboration, and include the quenching factor uncertainty as part of the combined systematic uncertainty $\sigma_\alpha = 16.9\%$, with a dedicated contribution of $7.3\%$~\cite{Lindner:2024eng,Chattaraj:2025fvx}.

\subsection{DUNE Near Detector}
\label{sec:analysis:dune}


DUNE is a next-generation long baseline neutrino oscillation experiment. Its ND complex, located at Fermilab, will be exposed to the high-intensity neutrino beam produced by the Long-Baseline Neutrino Facility (LBNF)~\cite{DUNE:2015lol}. The ND is equipped with a Liquid Argon Time Projection Chamber (LArTPC), with a fiducial mass about 67 tons and excellent particle identification capabilities. 
The DUNE ND design supports the Precision Reaction Independent Spectrum Measurement (PRISM) program, where the ND liquid Argon system can be moved transversely with respect to the beam to sample off-axis fluxes (up to 30\,m from the beam axis)~\cite{DUNE:2021tad}. 
For simplicity we consider conservatively only the on-axis (0\,m) configuration, which receives the highest neutrino fluxes and dominates the expected event rate. This simplification provides rather conservative sensitivities of DUNE ND, as inclusion of the off-axis data would definitely suppress the spectral uncertainties and improve the overall statistical power.

The DUNE neutrino beam is composed of $\nu_e$, $\bar{\nu}_e$, $\nu_\mu$, $\bar{\nu}_\mu$ with GeV energies under both the neutrino and anti-neutrino modes~\cite{DUNE:2015lol}. The corresponding on-axis neutrino fluxes can be found in Refs.~\cite{DUNEFluxesFields2017,DUNEFluxesGlaucus,Mathur:2021trm}. The total number of protons on target (POT) per year is expected to be $N_{\text{POT}} = 1.1 \times 10^{21}$~\cite{DUNE:2016hlj,DUNE:2021tad}. 
As discussed in Ref.~\cite{Ballett:2019bgd}, 
the electron detection threshold $T_e^{\rm min}$ can be as low as 30\,MeV~\cite{DUNE:2016ymp}.
So in the analysis we set the range of 
$T_e \geq T_e^{\rm min}$.


At the DUNE ND, the signal process produces a final-state electron that is experimentally indistinguishable from the elastic $\nu(\bar\nu)$-$e^-$ scattering. Therefore, the SM $\nu(\bar\nu)$-$e^-$ scattering constitutes the dominant irreducible background. The reducible backgrounds arise from the charged-current quasi-elastic (CCQE) neutrino scattering and mis-identified $\pi^0$ events~\cite{DeRomeri:2019kic,Melas:2023olz}, whose contributions can be effectively suppressed by the cut on the variable $E_e\theta_e^2$, where $E_e \equiv m_e + T_e$ and $\theta_e$ are the electron total energy and scattering angle, respectively. The kinematic relation is given by 
\begin{equation}
1 - \cos\theta_e = \frac{m_e}{E_e}(1 - y) \,,
\end{equation}
where the inelasticity $y \equiv T_e/E_\nu$ is
for the SM $\nu(\bar\nu)$-$e^-$ scattering.
For the inelastic scattering processes, the
inelasticity becomes~\cite{Candela:2024ljb} 
\begin{equation}
  y
\equiv
  \frac {T_e}{E_\nu}
  \left[1 + \frac{m_\chi^2}{2m_e T_e} \right] \,.
\end{equation}
The energy resolution of the LArTPC detector is estimated to be $\sigma_{E_e}/E_e \simeq 10\%/\sqrt{E_e/\text{GeV}}$, which has a negligible impact on our analysis~\cite{deGouvea:2019wav,Mathur:2021trm,Candela:2024ljb}. The angular resolution is taken to be $\sigma_\theta = 1^\circ$~\cite{DUNE:2021tad,deGouvea:2019wav}.

\begin{figure}[!t]
\centering
\includegraphics[width=0.48\linewidth]{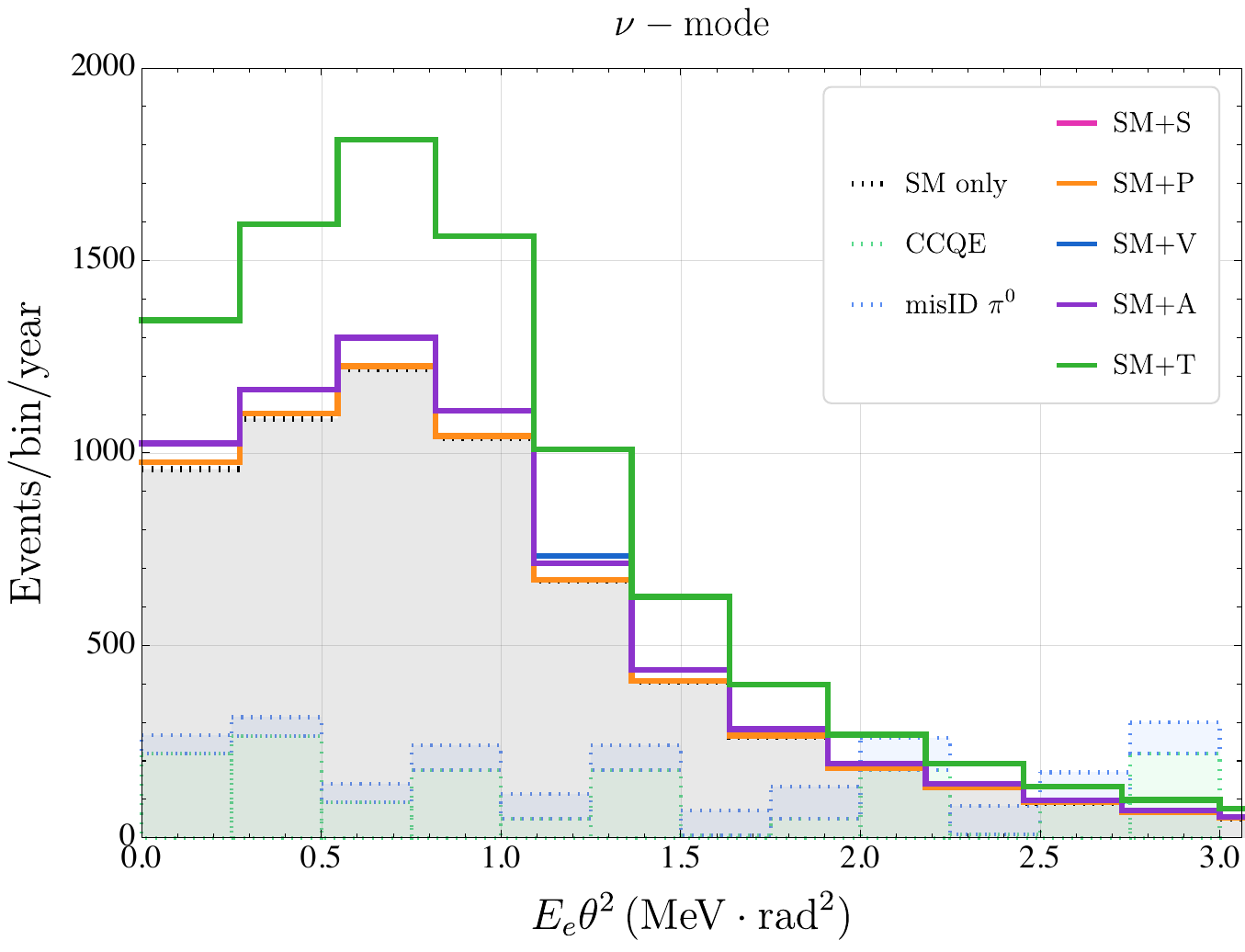}
\includegraphics[width=0.48\linewidth]{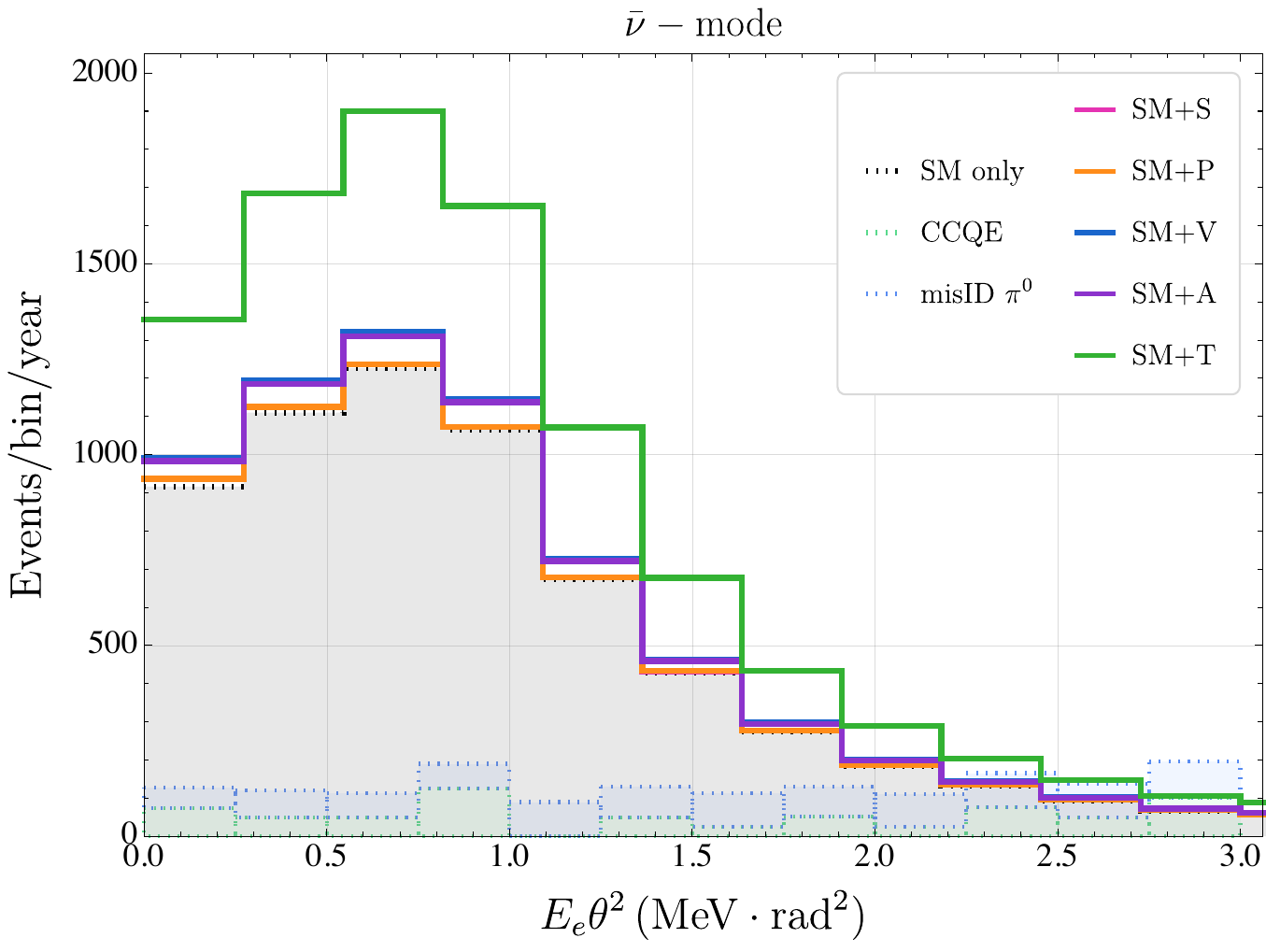}
\caption{Expected distributions of the variable
$x \equiv E_e\theta_e^2$ for the DUNE ND on-axis
configuration with one year of exposure, shown
separately for both the neutrino (left) and
antineutrino (right) modes. The shaded regions below
the dotted lines denote the SM background from
$\nu(\bar{\nu})$-$e^-$ elastic scattering and the
subdominant backgrounds from CCQE and mis-identified
$\pi^{0}$ events. The colored solid lines are for the
total events including the NP signal contribution for
the five Lorentz structures (scalar, pseudoscalar,
vector, axial-vector, and tensor). We have taken the
benchmark values of $m_\chi = 10\,\mathrm{MeV}$ and
$\Lambda_{i,e} = 650\,\mathrm{GeV}$ for all the NP
cases.}
\label{fig:DUNE eventnum}
\end{figure}

The number of events at the DUNE ND as a function of $E_e \theta_e^2$ can be calculated as~\cite{Candela:2024ljb}: 
\begin{equation}
 \frac{\mathrm{d}{\cal N}}{\mathrm{d} (E_e\theta_e^2)} \ = \ t_\mathrm{run} {N}_e N_\mathrm{POT} \sum_{\alpha} \int_{E_\nu^{\min}}^{E_\nu^{\max}}\mathrm{d}E_\nu\frac{\mathrm{d}\Phi_{\nu_\alpha}(E_\nu)}{\mathrm{d}E_\nu} \frac{\mathrm{d}\sigma_{\nu_\alpha}}{\mathrm{d} (E_e\theta_e^2)} \,,
\end{equation}
where the differential cross section is
\begin{equation}
\frac{\mathrm{d}\sigma_{\nu_\alpha}}{\mathrm{d}(E_e\theta_e^2)} \ = \ \left.\frac{E_\nu}{2m_e}\frac{\mathrm{d}\sigma_{\nu_\alpha}}{\mathrm{d}T_e}\right|_{T_e}.
\end{equation}
Here we have set $t_\text{run} = 7$ years (3.5 years in $\nu$ mode plus 3.5 years in $\bar{\nu}$ mode), $N_e$ is the number of electron targets in the LArTPC detector, ${\rm d}\Phi_{\nu_\alpha}/{\rm d}E_\nu$ is the (anti)neutrino flux for the different neutrino flavors $\nu_\alpha=\{\nu_e, \nu_{\mu}, \bar{\nu}_e, \bar{\nu}_\mu\}$, and $T_e^{\text{max}}$ is determined by kinematics. 
The differential distributions for the SM $\nu(\bar\nu)$-$e^-$ elastic scattering, the CCQE and mis-identified $\pi^0$ backgrounds, and the NP signal contributions are shown in \gfig{fig:DUNE eventnum} 
as functions of $x \equiv E_e\theta_e^2$. The left and right panels are for the neutrino and antineutrino modes, respectively. For all the NP scenarios with the five Lorentz structures, we have set the benchmark values of $m_\chi = 10$\,MeV and $\Lambda_{i,e} = 650$\,GeV. All spectra are normalized to one year of exposure and are grouped into 
12 bins in the range of $x\in[0,3]\,\mathrm{MeV\cdot rad^2}$.  
The SM $\nu(\bar\nu)$-$e^-$ elastic scattering cross sections used in our analysis follow the standard electroweak expressions, which can be found in e.g.  Ref.~\cite{Dev:2021xzd}.


The DUNE ND sensitivities are estimated by performing a $\chi^2$ analysis using 12 bins in $E_e\theta_e^2$~\cite{Melas:2023olz,Candela:2024ljb}. 
The $\chi^2$ function is defined as:
\begin{align}
\chi^2 = 2 \sum_{i=1}^{12} \left[ {\cal N}_{i}^{\text{th}} - {\cal N}_{i}^{\text{exp}} + {\cal N}_{i}^{\text{exp}} \log \left( \frac{{\cal N}_{i}^{\text{exp}}}{{\cal N}_{i}^{\text{th}}} \right) \right] + \left( \frac{\alpha_1}{\sigma_{\alpha_1}} \right)^2 + \left( \frac{\alpha_2}{\sigma_{\alpha_2}} \right)^2,
\end{align}
where $i$ is the index for the 12 bins, and
\begin{subequations}
\begin{align}
{\cal N}^{\text{th}} & \equiv (1 + \alpha_1) {\cal N}_{\text{SM}} + (1 + \alpha_2) {\cal N}_{\text{bkg}}+ {\cal N}_{\text{NP}} \,, \\
\label{eqn:N:exp}
{\cal N}^{\text{exp}} & \equiv {\cal N}_{\text{SM}} + {\cal N}_{\rm bkg} \,,
\end{align}
\end{subequations}
with ${\cal N}_{\rm SM,\, bkg,\, NP}$ the numbers of events for the SM $\nu(\bar\nu)$-$e^-$ scattering, the  CCQE and $\pi^0$ mis-identification backgrounds, and the NP contributions, respectively.
The nuisance parameters $\alpha_{1,\,2}$ are for
the flux ($\sigma_{\alpha_1} = 5\%$) and background ($\sigma_{\alpha_2} = 10\%$) normalization uncertainties, respectively~\cite{Candela:2024ljb}.
The expected DUNE ND sensitivities of $m_\chi$ and $\Lambda_{i,e}$ for the four-fermion interactions in \geqn{lagrangian} 
are derived by minimizing the $\chi^2$ function with respect to both the nuisance parameters $\alpha_i$ and the NP parameters.


\section{Hadronic absorption operators}
\label{sec:results:cevns}

We obtain in \gsec{sec:limits:cevns} the exclusion limits of COHERENT and CONUS+ on the effective cutoff scales $\Lambda_{i,N}$ for the couplings of neutrino and dark particles to nucleons, as functions of the dark particle mass $m_\chi$. In \gsec{sec:LHC:cevns}, we collect the other existing constraints on the cutoff scales $\Lambda_{i,N}$ that are comparable to or more stringent than the COHERENT and CONUS+ limits. For the sake of completeness, all other weaker limits are listed in \gsec{sec:weaker:cevns}. While the CE$\nu$NS experiments yield weaker constraints than the collider searches in the EFT (heavy mediator) limit, the CE$\nu$NS experiments can dominate over collider searches for certain regions of the parameter space in the presence of light mediators~\cite{Brdar:2018qqj,Chang:2020jwl,Chao:2021bvq,Candela:2023rvt,Candela:2024ljb,DeRomeri:2025csu}. Such light mediator scenarios are complementary to but not directly comparable with the EFT framework adopted in this work.


\subsection{Exclusion limits from COHERENT and CONUS+}
\label{sec:limits:cevns}



\begin{figure}[!t]
\centering
\includegraphics[width=0.48\linewidth]{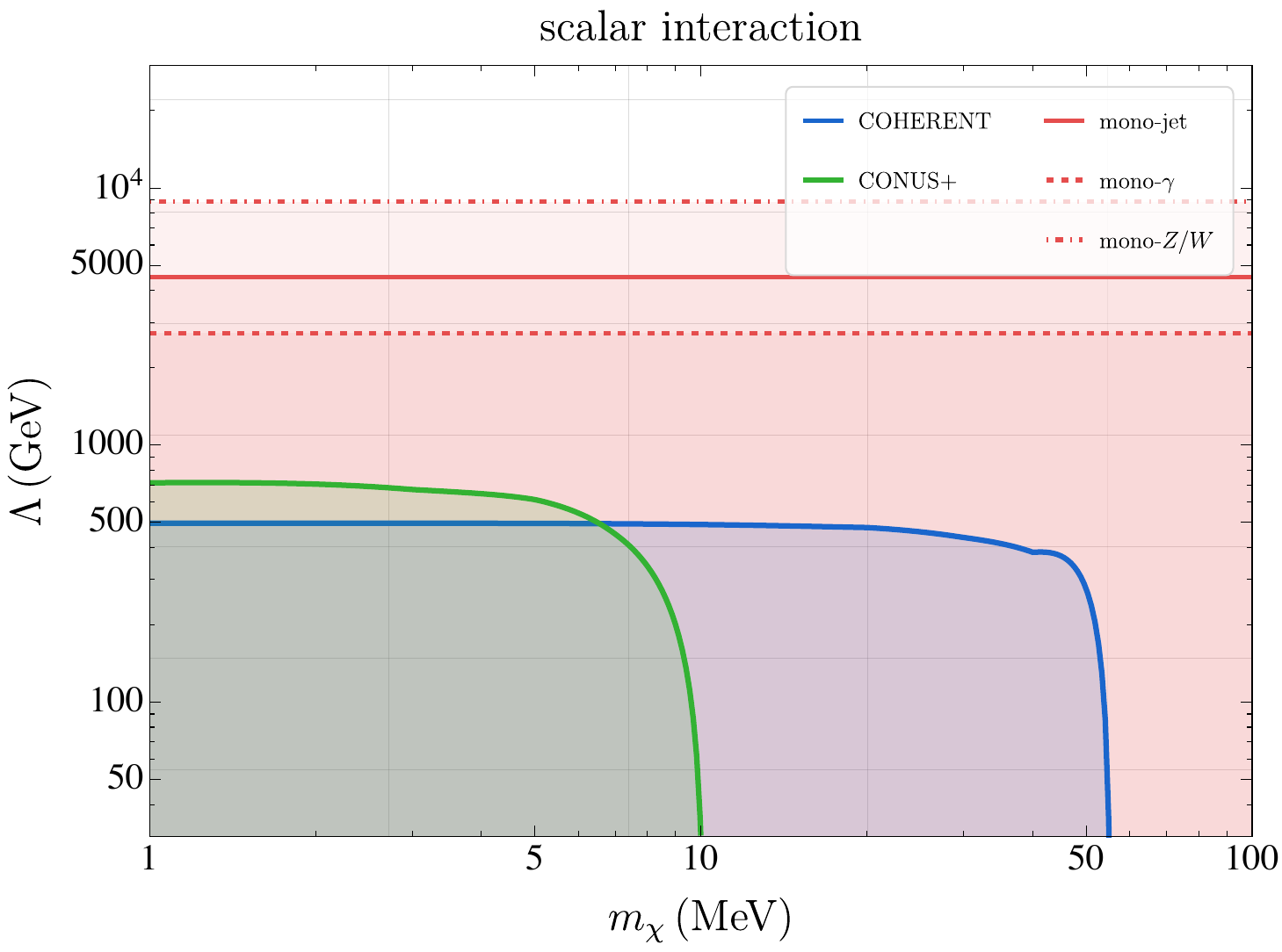}
\includegraphics[width=0.48\linewidth]{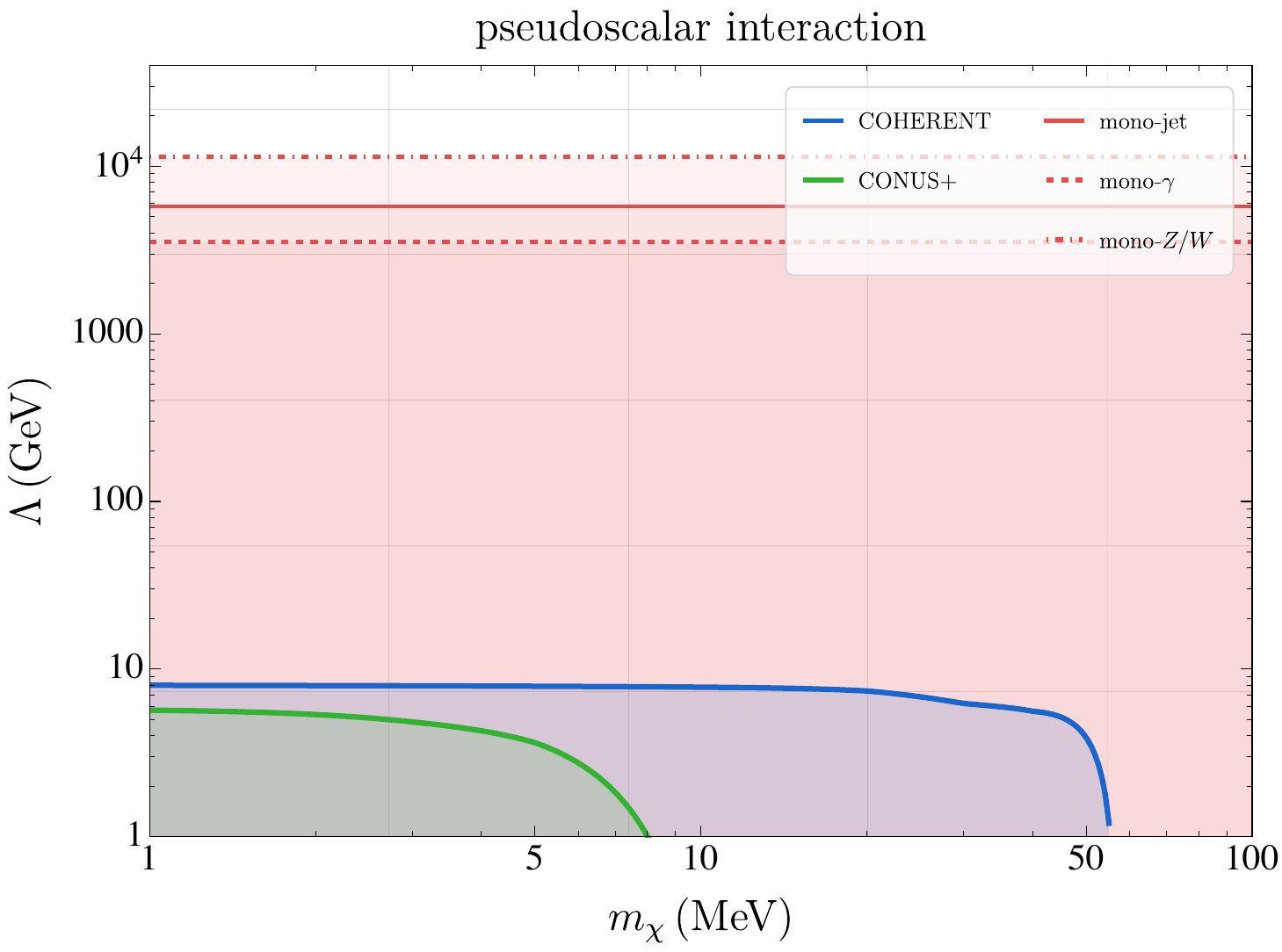}
\includegraphics[width=0.48\linewidth]{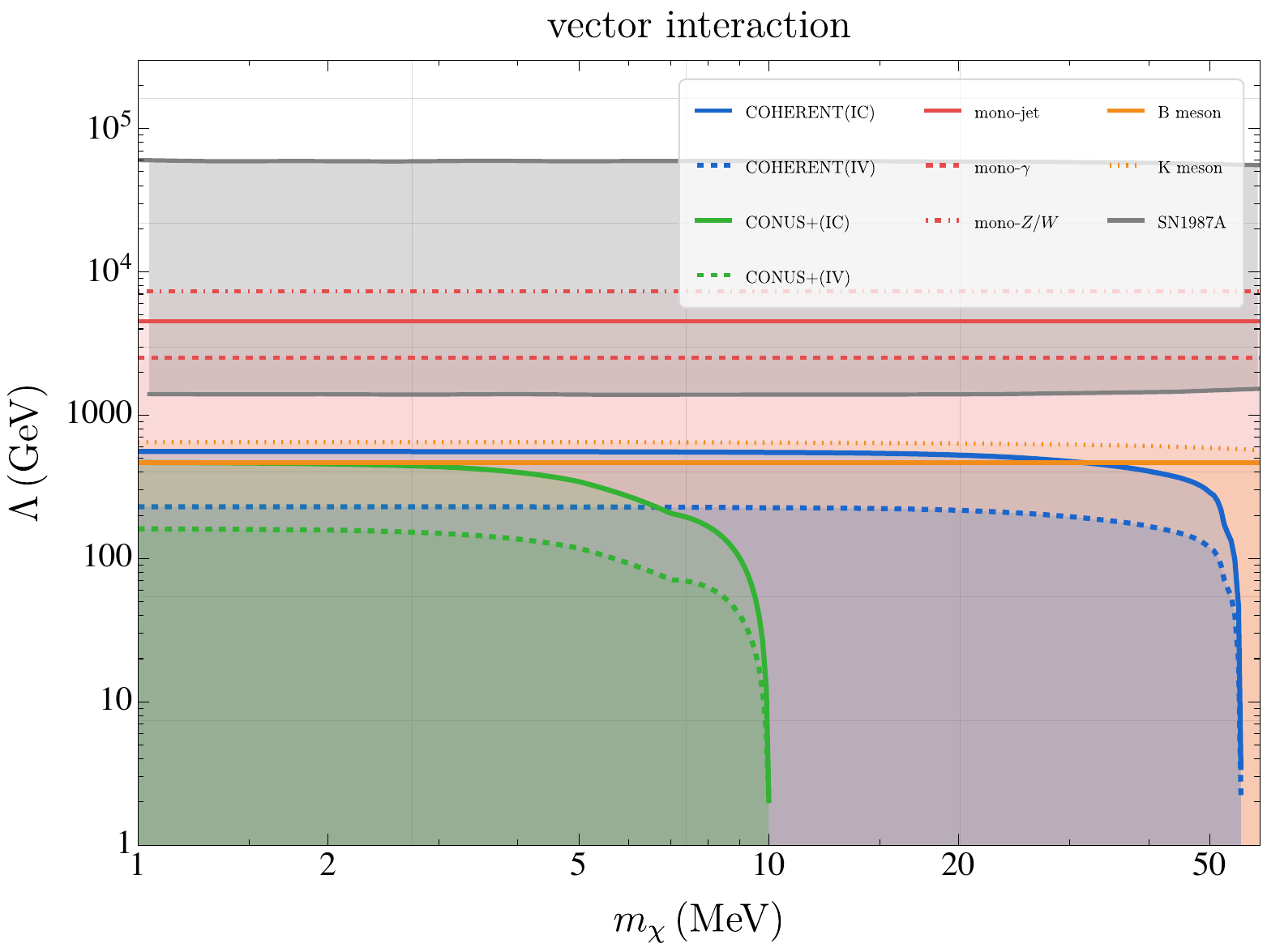}
\includegraphics[width=0.48\linewidth]{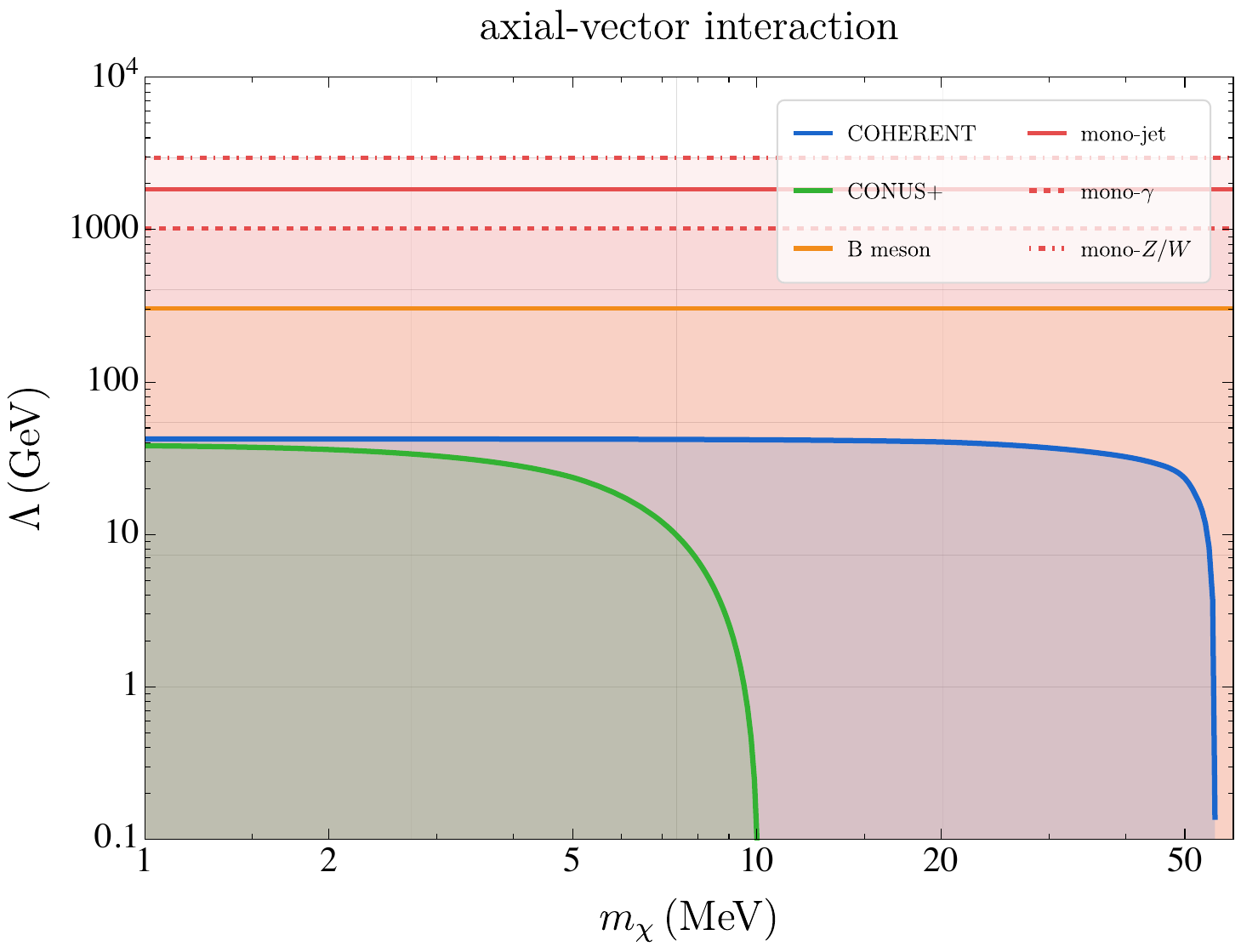}
\includegraphics[width=0.48\linewidth]{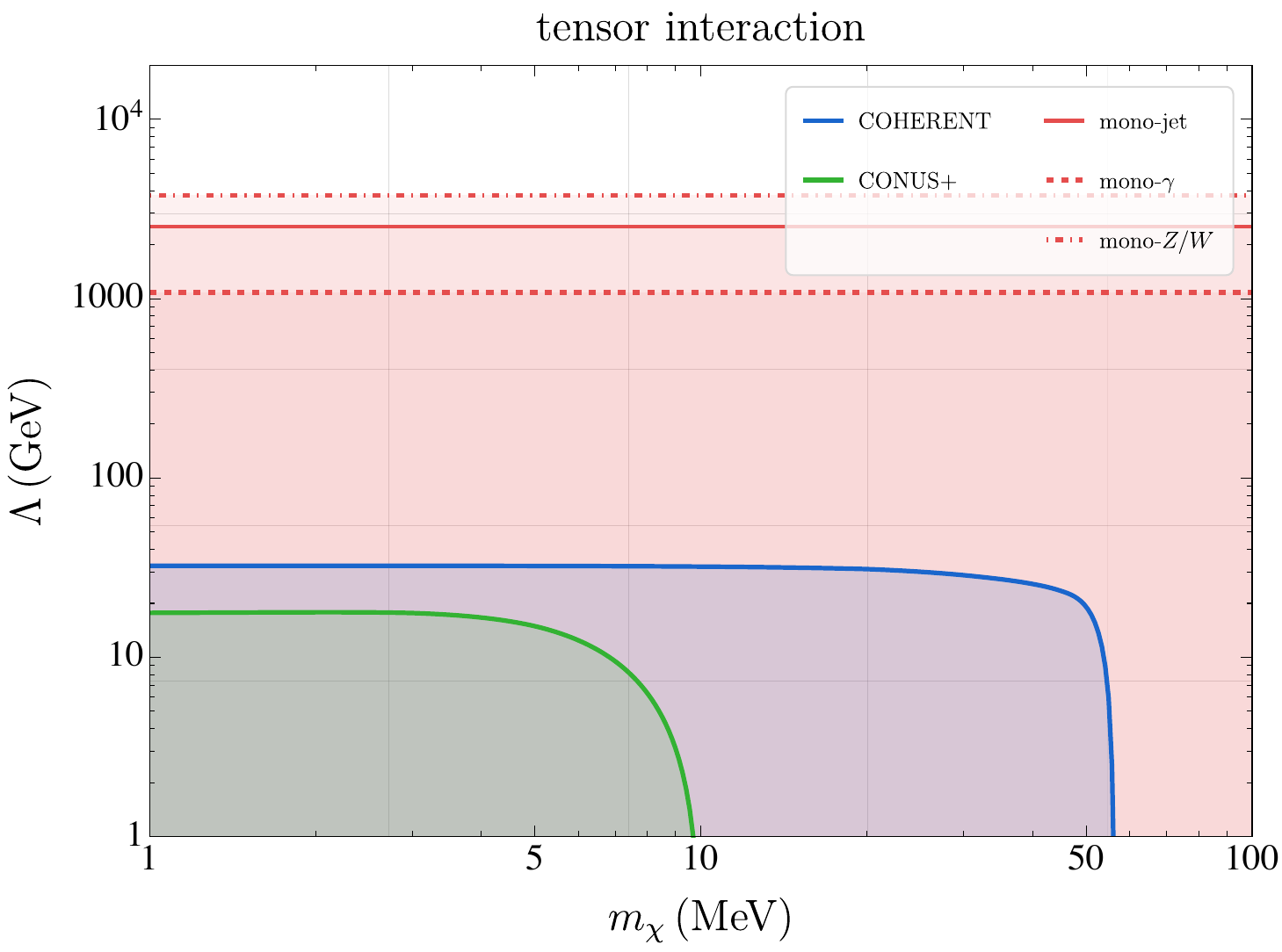}
\caption{The 90\%\,C.L. limits on the dark particle mass $m_\chi$ and the cutoff scales $\Lambda_{i,N}$ from the current COHERENT (shaded blue) and CONUS+ (shaded green) data, for all the five Lorentz structures of scalar (top left), pseudoscalar (top right), vector (middle left), axial-vector (middle right) and tensor (bottom) couplings. In the middle left panel, the limits for the IV vector couplings are indicated by the regions below the dashed blue and green lines. Also shown are  the limits from the LHC mono-jet, mono-$\gamma$ and mono-$Z/W$ data~\cite{Ma:2024tkt}, the FCNC $B$ and $K$ meson decays~\cite{Liu:2025lbw} and the SN1987A observations~\cite{Lin:2025mez}.}
\label{fig:CEvNS sensitivity}
\end{figure}

Based on the analysis in \gsec{sec:analysis}, the excluded regions by the current COHERENT and CONUS+ data are shown as the blue and green shaded regions in \gfig{fig:CEvNS sensitivity}. The five panels are for the scalar, pseudoscalar, vector, axial-vector and tensor coupling forms, respectively. 
As stated in \gsec{sec:kinematics}, the dark particle mass ranges accessible in CE$\nu$NS experiments are primarily determined by the energies of the neutrino sources employed. For the CONUS+ experiment, which use reactor antineutrinos with typical energies of a few MeV, the exclusion limits extend only up to $m_\chi \lesssim \mathcal{O}(10\,\mathrm{MeV})$. In comparison, the COHERENT experiment is based on a spallation neutron source, producing neutrinos with energies up to several tens of MeV. Consequently, the excluded dark particle mass range can be extended to $m_\chi \lesssim \mathcal{O}(50\,\mathrm{MeV})$.


Let us now collect the exclusion limits on the cutoff scales $\Lambda_{i,N}$ for the different Lorentz structures of the effective interactions.
\begin{itemize}
    \item For the scalar interaction, both COHERENT and CONUS+ benefit from the coherent enhancement associated with the spin-independent coupling in the neutrino-nucleus scattering process. The exclusion limit from COHERENT reaches $\Lambda_{S,N} \simeq 450\,\mathrm{GeV}$, while CONUS+, due to its lower recoil energy threshold and operation deep in the fully coherent regime, provides a stronger constraint of $\Lambda_{S,N} \simeq 692\,\mathrm{GeV}$, as seen in the top left panel of \gfig{fig:CEvNS sensitivity}.

    \item In contrast, the pseudoscalar interaction has the strongest suppression in CE$\nu$NS experiments, as presented in the top-right panel of \gfig{fig:CEvNS sensitivity}. The corresponding exclusion limits are only $\Lambda_{P,N} \simeq 8\,\mathrm{GeV}$ for COHERENT and $\Lambda_{P,N} \simeq 6\,\mathrm{GeV}$ for CONUS+, nearly two orders of magnitude weaker than those for scalar interactions. This severe reduction originates from both the momentum and spin suppressions that are inherent in the non-relativistic nuclear scatterings for a pseudoscalar coupling.

    \item For the vector coupling, the exclusion limits in the IC scenario are comparable to those of scalar interactions. COHERENT and CONUS+ yield exclusion limits of $\Lambda_{V,N}^{\mathrm{IC}} \simeq 486\,\mathrm{GeV}$ and $\Lambda_{V,N}^{\mathrm{IC}} \simeq 388\,\mathrm{GeV}$, respectively, as indicated by the dashed blue and green lines in the middle-left panel of \gfig{fig:CEvNS sensitivity}. This reflects the coherent enhancement of vector currents in nuclear targets. 
    
    In the IV scenario with $\Lambda_{V,\,p} = -\Lambda_{V,\,n}$, the effective nuclear coupling scales with neutron excess $({\sf Z} - {\sf N})$, leading to severe suppression. The exclusion limits decrease to $\Lambda_{V,N}^{\mathrm{IV}} \simeq 200\,\mathrm{GeV}$ for COHERENT and $\Lambda_{V,N}^{\mathrm{IV}} \simeq 133\,\mathrm{GeV}$ for CONUS$+$, respectively, which are depicted as the dotted blue and green lines in the middle-left panel of \gfig{fig:CEvNS sensitivity}. This behavior demonstrates the strong dependence of CE$\nu$NS constraints on the nuclear composition of the detector material in isospin relevant scenarios.

    \item The axial-vector and tensor interactions are also to some extent suppressed due to their dependence on the nuclear spin, as seen in the middle right and bottom panels of \gfig{fig:CEvNS sensitivity}. The exclusion limits obtained by COHERENT are $\Lambda_{A,N} \simeq 44\,\mathrm{GeV}$ and $\Lambda_{T,N} \simeq 25\,\mathrm{GeV}$, while CONUS+ yields $\Lambda_{A,N} \simeq 38\,\mathrm{GeV}$ and $\Lambda_{T,N} \simeq 18\,\mathrm{GeV}$, respectively. These results demonstrate the intrinsic disadvantage of spin-dependent interactions in CE$\nu$NS experiments dominated by nuclei with small spins. 
\end{itemize} 
As clearly seen in the panels of \gfig{fig:CEvNS sensitivity}, the CE$\nu$NS experiments exhibit a clear hierarchy among different Lorentz structures: the scalar and IC vector interactions have the strongest exclusion limits, whereas the pseudoscalar, axial-vector, and tensor interactions are significantly suppressed. As we will see in \gsec{sec:result:DUNE}, this pattern stands in sharp contrast to the results provided by the neutrino-electron scattering at DUNE ND, underscoring the complementarity between these two classes of experiments.

It should be emphasized that the parameter space probed by COHERENT and CONUS+ has already been constrained by the existing LHC 13\,TeV data~\cite{Ma:2024tkt}. The constraints on the cutoff scales $\Lambda_{i,N}$ can be significantly improved by the future COHERENT and CONUS+ data, which however, are still below the current LHC limits~\cite{Chang:2020jwl} to be detailed below.
As a result, the CE$\nu$NS bounds presented here do not constitute the primary discovery reach of this work, but instead serve as important and realistic benchmarks that provide complementary information to the DUNE ND results.


\subsection{Limits from LHC data, FCNC meson decays and SN1987A}
\label{sec:LHC:cevns}

The leading constraints on the effective interactions in \geqn{lagrangian} for nucleon are from the direct production of dark particles at the LHC, the FCNC meson decays, and the supernova explosion.

\paragraph{LHC limits} 
The effective operators involving nucleons are severely constrained by the existing LHC data. To adopt the LHC constraints, we need to correlate the effective couplings of nucleons in \geqn{lagrangian} to that of quarks:
\begin{equation}
\label{eqn:Lagrangian:quark}
\mathcal{L}_{\mathrm{eff},q}=\sum_{i} \frac{1}{\Lambda_{i,\,q}^2}\left(\bar{\chi}\right.\Gamma^i P_L\left.\nu\right)\left(\bar{q}\operatorname{\Gamma}_i q\right) ~+~ \mathrm{h.c.} \,,
\end{equation}
with $\Lambda_{i,q}$ the corresponding cutoff scales. As the couplings involving gluons are always at the 1-loop or higher order, for simplicity we consider here only the couplings with quarks. The cutoff scales $\Lambda_{i,\,q}$ are correlated with  $\Lambda_{i,\,{\cal N}}$ through the form factors ${\cal F}_i$ via the relation~\cite{Shifman:1978zn,Drees:1993bu,Crivellin:2013ipa,Bishara:2017pfq,Ma:2024tkt}
\begin{equation}
{\cal F}_i\left(\bar{N}\operatorname{\Gamma}_i N\right) \simeq 
\left(\bar{q}\operatorname{\Gamma}_i q\right).
\end{equation}
This implies that 
\begin{equation}
\Lambda_{i,N}^2 \simeq {\cal F}_i \, \Lambda_{i,q}^2 \,,\quad \text{with} \quad i=S,P,V,A,T \,. 
\end{equation}
More details can be found in \gapp{app:nucleon}.

Neutrinos, DM and other invisible particles can be produced at the high energy hadron colliders, e.g. from their interactions with quarks, and emerge as missing transverse energy $\slashed{E}_T$ at the detectors~\cite{Kahlhoefer:2017dnp,Boveia:2018yeb,Krnjaic:2022ozp}. The main LHC constraints are from the processes of $q \bar{q} \to X+ \nu \bar\chi \, (\bar\nu \chi)$, with $X = $jet, $\gamma$ and $Z/W$. This corresponds to the mono-jet~\cite{ATLAS:2021kxv,CMS:2021far}, mono-$\gamma$~\cite{ATLAS:2020uiq} and mono-$Z/W$~\cite{ATLAS:2021gcn,ATLAS:2018nda,ATLAS:2024rlu} signals at the LHC, respectively. All these analysis are based on the simplified DM models with a heavy mediator. These LHC limits have been re-interpreted and recast onto the EFT couplings of quarks in \geqn{eqn:Lagrangian:quark}~\cite{Ma:2024tkt}. The resultant constraints on the cutoff scales $\Lambda_{i,N}$ at the nucleon level are presented as the red lines with shaded regions in \gfig{fig:CEvNS sensitivity}. It is very clear that for all the five Lorentz structures the current LHC limits are much more stringent than the current constraints from COHERENT and CONUS+.




\begin{table}[!t]
\centering
\caption{The 90\%\,C.L. limits on the cutoff scales $\Lambda_{i,N}$ in Eq.\,(\ref{lagrangian}) from the current COHERENT and CONUS+ data. Also shown are the existing limits from the LHC mono-jet, mono-$\gamma$ and mono-$Z/W$ data, FCNC meson decay and SN1987A observations. All the limits are for the mass of $m_\chi \to 0$, and in unit of GeV.}
\label{tab:cevns}
\vspace{5pt}
\scalebox{0.86}{
\begin{tabular}{c|c|cc|ccc|cc|c}
\toprule
\multicolumn{2}{c|}{\multirow{2}{*}{int.}} & COHE- & \multirow{2}{*}{CONUS+} & \multicolumn{3}{c}{LHC~\cite{Ma:2024tkt}}& \multicolumn{2}{|c|}{meson~\cite{Liu:2025lbw}} & \multirow{2}{*}{SN1987A~\cite{Lin:2025mez}} \\
\cline{5-9}
\multicolumn{2}{c|}{} & RENT & & mono-jet & mono-$\gamma$ & mono-$Z/W$ & $B$ & $K$ & \\
\midrule
\multicolumn{2}{c|}{$S$} & 450 & 710 & 4.5$\times 10^3$ & 2.7$\times 10^3$ & 8.8$\times 10^3$ & $-$ & $-$ & $-$  \\ \hline
\multicolumn{2}{c|}{$P$} & 8 & 6 & 5.8$\times 10^3$ & 3.5$\times 10^3$ & 1.1$\times 10^4$ & $-$ & $-$ & $-$  \\ \hline
\multirow{2}{*}{$V$} & IC & 486 & 388 & 4.5$\times 10^3$ & 2.5$\times 10^3$ & 7.3$\times 10^3$ & 467 & 646 & [$1.5\times10^3$, $6.0\times10^4$] \\ \cline{2-10}
& IV & 200 & 133 & $-$ & $-$ & $-$ & $-$ & $-$ & $-$ \\ \hline
\multicolumn{2}{c|}{$A$} & 44 & 38 & 1.8$\times 10^3$ & 1.0$\times 10^3$ & 2.9$\times 10^3$ & 304 & $-$ & $-$ \\ \hline
\multicolumn{2}{c|}{$T$} & 25 & 18 & 2.5$\times 10^3$ & 1.1$\times 10^3$ & 3.8$\times 10^3$ & $-$ & $-$ & $-$ \\
\bottomrule
\end{tabular}}
\end{table}

\paragraph{FCNC meson decay limits}

The couplings of DM $\chi$ with quarks in the form of\begin{equation}
\label{eqn:EFT:quark-DM}
 (\bar{\chi} \Gamma^i \chi) (\bar{q} \Gamma_i q),
\end{equation}
could induce the exotic FCNC decays, e.g. $d_j \to d_i \chi\bar\chi$. Such processes are constrained by the precise data of the meson decays $B \to K^{(\ast)} \nu \bar\nu$, $B \to \pi \nu \bar\nu,\, \rho \nu \bar\nu$ and $K \to \pi \nu \bar\nu$~\cite{Fayet:2006sp,McKeen:2009rm,Kamenik:2011vy,Darme:2020ral,Costa:2022pxv,Liu:2025lbw,Mescia:2026xju}. Here we adopt the recent FCNC limits from Ref.~\cite{Liu:2025lbw}. When we recast these limits onto the channels $d_j \to d_i \nu \bar\chi$ induced by the couplings in Eq.~(\ref{lagrangian}), the corresponding limits are in principle different from the limits on $d_j \to d_i \chi \bar\chi$, depending on the mass $m_\chi$. For the mass range of $m_\chi \lesssim 100$ MeV of interest in this paper, which is much smaller than the Kaon and $B$ meson masses, the difference is insignificant and can be safely neglected. Constraints are derived on the products of vector and axial-vector quark and dark sector couplings $g_q^V g_\chi^V$ and $g_q^A g_\chi^A$ from the rare $B$ and $K$ decays. For the heavy gauge boson $Z'$ mediator with mass $m_{Z'}$, these bounds can be mapped onto the quark-level four-fermion operators in \geqn{eqn:EFT:quark-DM} with the corresponding effective quark-level cutoff scales $\Lambda_{i,q} \equiv m_{Z'}/\sqrt{g_f^i g_\chi^i}$ for $i=V,\,A$. 
In $B$ meson decays, the channels $B^0 \to K^0 \nu \bar\nu,\, \pi^0 \nu \bar\nu$ and $B^0 \to K^{\ast0} \nu \bar\nu$ provide the most stringent constraints
for the vector and axial-vector couplings, respectively. 
The kaon decay $K^+ \to \pi^+ \nu \bar\nu$ offers comparable constraints on the effective vector coupling. 
Following the same procedure above, the limits on the quark-level couplings can be converted onto the constraints on the cutoff scales $\Lambda_{i,N}$ at the nucleon level. As seen in the middle-left and middle-right panels of \gfig{fig:CEvNS sensitivity}, the $B$ meson decays lead to the limits of $\Lambda_{V,N}^{\rm IC}\gtrsim 467$\,GeV and $\Lambda_{A,N} \gtrsim304$\,GeV, while the $K$ meson constraint is $\Lambda_{V,N}^{\rm IC} \gtrsim 646$\,GeV.




\paragraph{Supernova limits} 

The scattering process in \geqn{eqn:process:nuN} could also occur in the supernova core. Therefore, the cutoff scales $\Lambda_{i,N}$ are tightly constrained by the supernova observations. In particular, copious of $\chi$ can be produced in the $\nu(\bar\nu)$-nucleon scattering in supernovae, and then escape from the stars. By requiring that the energy carried away by dark particles does not exceed a certain fraction of the total neutrino emission (the Raffelt cooling criterion)~\cite{Raffelt:1996wa}, very stringent constraints can be obtained for the mass range of $m_{\chi} \lesssim {\cal O}(100 \, {\rm MeV})$. The supernova cooling constraint on the IC vector coupling case have been recently obtained in Ref.~\cite{Lin:2025mez}. Although the supernova limit in this paper is based on the assumption of $\chi$ being DM, it applies also to our case of $\chi$ being unstable dark particle instead of DM, as the decay length of $\chi$ is much larger than the typical size of supernova cores. 
It turns out the range of 1.5\,TeV$ \lesssim \Lambda_{V,N}^{\rm IC} \lesssim$ 60\,TeV is excluded by the SN1987A data, which is presented as the gray shaded band in \gfig{fig:CEvNS sensitivity}.

For the convenience of readers, the limits on the cutoff scales $\Lambda_{i,N}$ from the current COHERENT and CONUS+ data, and the existing LHC, meson and supernova limits above have been compiled
together in \gtab{tab:cevns}. All these limits are for the light dark particle with vanishing mass $m_\chi \to 0$. It is transparent in this table that the current most stringent constraints are from the high-energy LHC data and the SN1987A observations.

\subsection{Other weak constraints}
\label{sec:weaker:cevns}

In this subsection, we collect all the weak and inapplicable constraints on the effective couplings of neutrino and dark particle $\chi$ with nucleons in \geqn{lagrangian}.
\begin{itemize}
    \item \textbf{(Semi-)invisible quarkonium decays}. The  searches of (semi-)invisible quarkonium decays at BaBar provide direct bounds on the quark-level process $q \bar{q} \to \chi\bar\chi$ and $q \bar{q} \to \chi\bar\chi\gamma$, which are complementary to the mono-$X$ constraints at the LHC~\cite{Fernandez:2014eja}.
In particular, the BaBar collaboration reports the upper limits $\mathcal{B}(\Upsilon(1S)\to \text{invisible})<3\times 10^{-4}$ and
$\mathcal{B}(\Upsilon(3S)\to \gamma+\text{invisible})<(0.7 \sim 31)\times 10^{-6}$, depending on the missing mass window. As in the case of FCNC meson decays in \gsec{sec:LHC:cevns}, these data can be used to constrain the channels of $q\bar{q} \to \nu \bar\chi,\, \nu\bar\chi\gamma$, which can arise from the couplings of neutrino and dark particle $\chi$ with the bottom quark. It turns out that $\Lambda \gtrsim \mathcal{O} (100)\,\mathrm{GeV}$.

\item \textbf{Radiative $Z$ and tauon decays}. The couplings in \geqn{lagrangian} contribute at the 1-loop order to the invisible decay of $Z$ boson via the channels of $Z \to \nu \bar\chi,\,\chi\bar\nu$, which is quite similar to the effects of neutrino NSIs~\cite{Davidson:2003ha}. The corresponding limits on the cutoff scales are $\Lambda \gtrsim 200$\,GeV. 
The couplings of neutrinos (and the dark particle $\chi$) with quarks induce extra 1-loop contributions to the decay of tauon $\tau \to \pi \nu (\pi \chi)$. However, limited by the precision of tauon data, the corresponding constraints are expected to be weaker~\cite{Davidson:2003ha}.    

\item 
\textbf{DM direct detection}. 
The DM direct detection experiments provide an important platform for probing neutrino relevant NP. If $\chi$ is identified as the DM particle, the effective coupling in \geqn{lagrangian} could induce the absorption of DM in the direct detection experiments via the process of $\chi + {\cal N} \to {\cal N} + \nu$. Thanks to the low energy thresholds and large exposures of the DM direct detection experiments, the corresponding limits on the cutoff scales $\Lambda_{i,N}$ can reach $\mathcal{O}(\mathrm{TeV})$ for the scalar and vector couplings
\cite{PandaX:2022osq,Majorana:2022gtu,CDEX:2022rxz,PICO:2025rku}. However, such constraints do {\it not} apply to our case, with $\chi$ being an unstable particle. 

\item 
\textbf{Neutrino signals in DM direct detection experiments}. The recent indications of CE$\nu$NS from solar $^8$B 
neutrinos at the PandaX-4T~\cite{PandaX:2024muv} and XENONnT~\cite{XENON:2024ijk} experiments have opened new opportunities to constrain  neutrino NSIs at very low momentum transfers. The analyses in Refs.~\cite{Blanco-Mas:2024ale,DeRomeri:2024iaw} are performed for the scattering of solar neutrinos with nuclei or electrons in the light mediator framework, and can not be directly compared with our EFT results.

\item \textbf{Light mediator portal dark particle searches}. The dark particle $\chi$ can couple to the SM particles via a light mediator, e.g. a light scalar or a dark photon at or below the GeV. Then the light mediator can be produced from meson decays or at the fixed target experiments, and then decays into a pair of dark particles~\cite{Batell:2018fqo,Alvey:2019zaa,Flambaum:2020xxo,Su:2022wpj,PandaX:2023tfq,BESIII:2026qsu,Batell:2021aja,Batell:2021blf,deNiverville:2012ij,Diamond:2013oda,Izaguirre:2013uxa,Batell:2014yra,Asai:2022zxw,Asai:2023mzl}. However, the constraints from such processes can not be used to directly constrain the effective four-fermion couplings in \geqn{lagrangian}. The dark particle $\chi$ can be pairly produced from the fixed target experiments, but the corresponding constraints are expected to be much weaker than the light mediator case. 

\item 
\textbf{Loop correction to CE$\nu$NS processes}. The operators involving $\chi$ could contribute at the 1-loop level to the CE$\nu$NS processes. As a result, the cutoff scales are constrained by the existing CE$\nu$NS data. The loop effects of the $\chi$ couplings with neutrino and nuclei via a light mediator and the corresponding COHERENT constraints are studied in Ref.~\cite{Chao:2021bvq}. However, suppressed by the loop factor, these limits are to some extent weaker.

\item \textbf{Supernova observations}. 
The radiative decays of axion-like particles may generate photon signals from supernovae~\cite{Caputo:2022mah,Fiorillo:2025yzf,Fiorillo:2025sln,Diamond:2023scc,Candon:2025ypl,Ferreira:2025qui,Ferreira:2022xlw,Calore:2021klc,Croon:2020lrf} and neutron stars~\cite{Diamond:2023cto,Dev:2023hax}. The Majoron-like particles, which are assumed to couple predominantly to neutrinos, could also induce rich neutrino signatures from supernovae~\cite{Choi:1987sd,Aharonov:1989ik,Kachelriess:2000qc,Farzan:2002wx,Fiorillo:2022cdq}. However, these limits 
can not be applied directly to constrain the radiative couplings of dark particle $\chi$ to neutrino and photon(s)~\cite{Dror:2020czw,Ge:2022ius}, as the production mechanism and decay products of $\chi$ in the stars are very different from the axion- and Majoron-like particles.  
\end{itemize}

\section{Leptonic absorption operators}
\label{sec:result:DUNE}

This section focuses mostly on the couplings of neutrinos and dark particle with electrons. The expected sensitivities at the DUNE ND are obtained in \gsec{sec:sensitivity:DUNE}. The relevant constraints from the current CHARM II and LEP data and the prospects at future high-energy lepton colliders are given in \gsec{sec:limits:LEP}. Some other weaker constraints are collected in \gsec{sec:limits:EnuES:weaker}.

\subsection{The sensitivities at DUNE ND}
\label{sec:sensitivity:DUNE}



\begin{figure}[!t]
    \centering
    
    \includegraphics[width=0.6\linewidth]{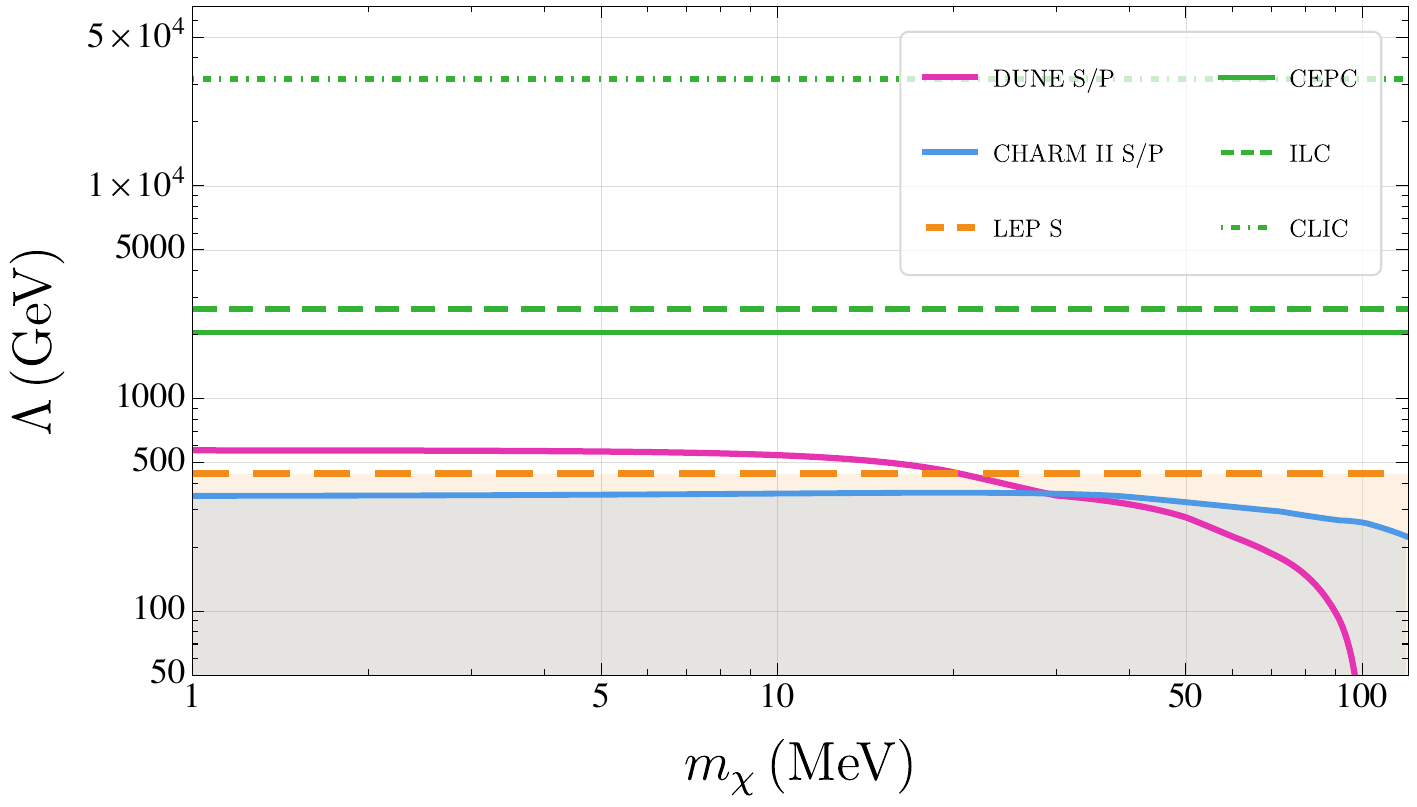}
    \includegraphics[width=0.6\linewidth]{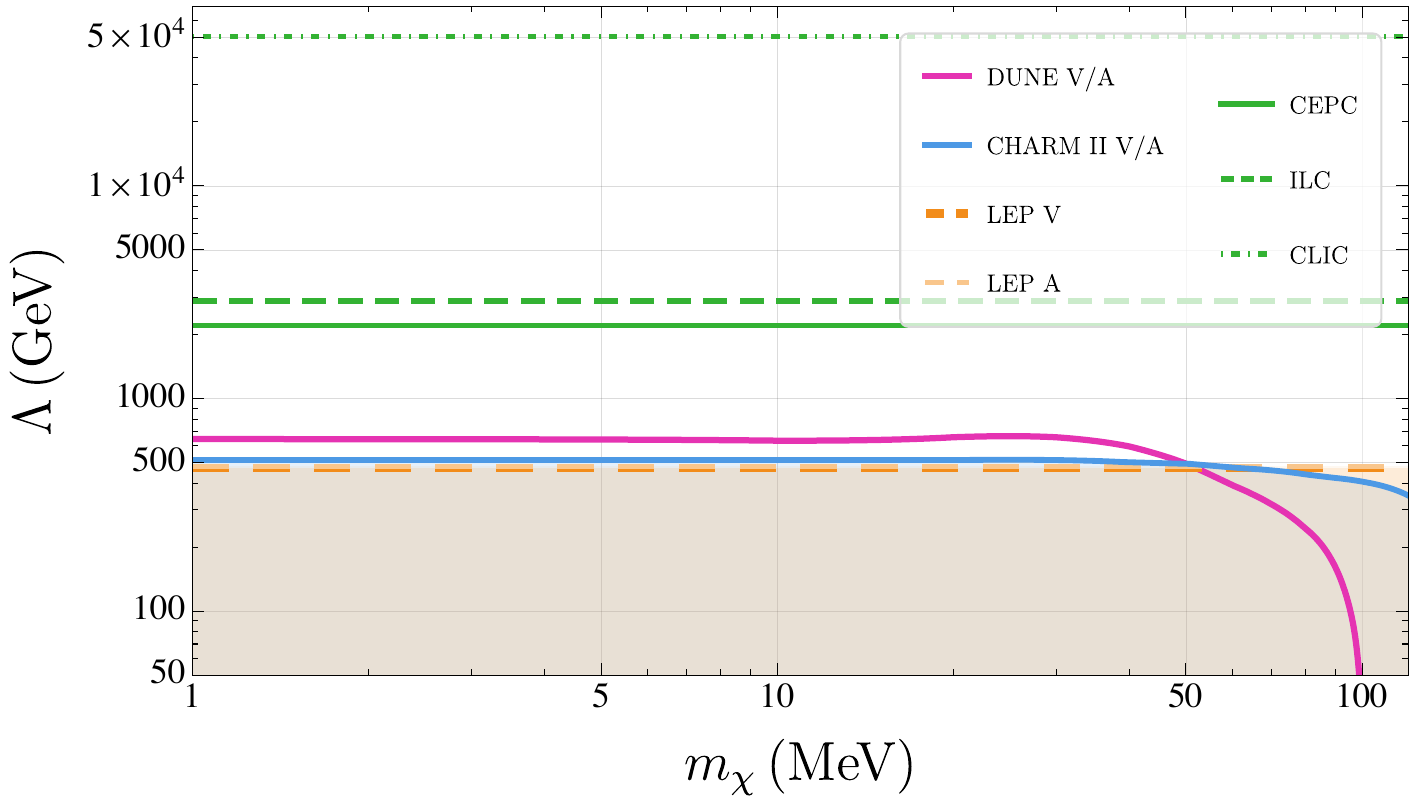}
    \includegraphics[width=0.6\linewidth]{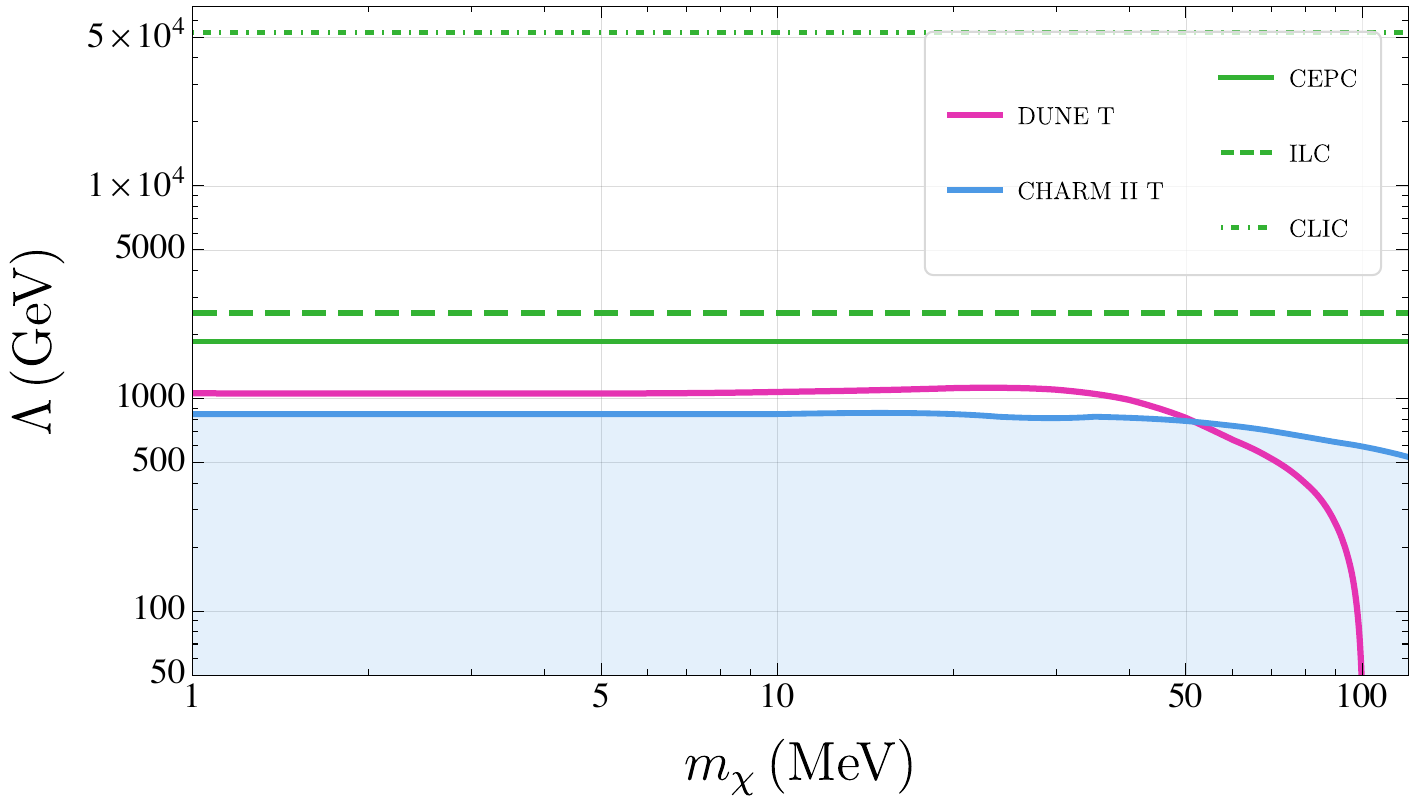}
    \caption{The 90\% C.L. sensitivities of DUNE ND to the dark particle mass $m_\chi$ and the cutoff scales $\Lambda_{i,e}$, for the scalar/pseudoscalar (top), vector/axial-vector (middle), and tensor (bottom) interactions, shown as the solid magenta curves.  Also shown are the existing constraints from CHARM II~\cite{Chen:2021uuw} and LEP~\cite{Fox:2011fx}, which are depicted as the shaded regions below the solid blue lines and dashed oranges lines, respectively. The green lines correspond to the projected sensitivities of future lepton colliders~\cite{Ge:2023wye}.
    See text for more details.}
    \label{fig:DUNE sensitivity}
\end{figure}

The typical energy of the neutrino beam in DUNE is at the GeV scale. As stated in \gsec{sec:kinematics}, the mass $m_\chi$ can be probed up to ${\cal O}(100\,{\rm MeV})$ at the DUNE ND, significantly higher than that of low-energy neutrino experiments based on reactors or spallation sources. 
Furthermore, comparing with the neutrino-nucleus scattering, the neutrino-electron scattering does not involve nuclear structure effects and there is no additional suppression caused by the nuclear spin or isospin structure. 
Benefiting from the high POT of DUNE and the well controlled background environment, the sensitivities of $\Lambda_{i,e}$ at the DUNE ND are comparable to or significantly higher than the limits on $\Lambda_{i,N}$ from COHERENT and CONUS+ in \gsec{sec:results:cevns}, depending on the Lorentz structures involved. Based on the analysis in \gsec{sec:analysis:dune}, the resultant sensitivity of DUNE ND to the effective cutoff scales $\Lambda_{i,e}$ are shown as the solid magenta lines in \gfig{fig:DUNE sensitivity}, as functions of $m_\chi$. 
The top, middle and bottom panels are for the scalar (pseudoscalar), vector (axial-vector) and tensor couplings, respectively. All sensitivities are shown at the 90$\%$\,C.L..
\begin{itemize}
    \item At DUNE ND, the recoil electrons in the (anti)neutrino-electron scattering are typically relativistic. Consequently, unlike in low-energy CE$\nu$NS measurements such as COHERENT and CONUS+ experiments, there is no suppression behavior of the pseudoscalar interaction. In particular, the projected sensitivities for the scalar and pseudoscalar cases almost coincide over the entire dark particle mass range, and are represented by the same solid magenta curve in the top panel of \gfig{fig:DUNE sensitivity}, reaching $\Lambda_{S(P),e}\simeq 560\,\rm{GeV}$ up to 560\,GeV. 
    
    \item Similarly, the sensitivities for the vector and axial-vector interactions are nearly identical, with $\Lambda_{V(A),e}\simeq 650\,\rm{GeV}$, as shown in the middle panel of \gfig{fig:DUNE sensitivity}. This indicates that in the energy range and detection channels of DUNE ND considered here, these operators lead to essentially the same sensitivity. 
    
    \item As seen in the bottom panel of \gfig{fig:DUNE sensitivity}, the tensor interaction gives the strongest sensitivity throughout the mass range, with the corresponding effective cutoff scale reaching $\Lambda_{T,e}\simeq 1.07\,\rm TeV$. 
    
    \item When the dark particle mass approaches the kinematic threshold, $m_\chi\sim 100\,{\rm MeV}$, the sensitivities for all the operators decrease rapidly due to kinematics.    
    \end{itemize}
For the convenience of readers, the sensitivities to $\Lambda_{i,e}$ in the limit $m_\chi\to0$ for all the five Lorentz structures at DUNE ND are collected in \gtab{tab:DUNE}.






\begin{table}[!t]
\centering
\caption{Expected sensitivity of the cutoff scales $\Lambda_{i,e}$ in Eq.\,(\ref{lagrangian}) at the 90\% C.L. for the DUNE ND. Also shown are the current CHARM II and LEP limits and prospects at future high-energy lepton colliders. The limits and prospects are for the dark particle mass $m_\chi \to 0$, and are given in GeV. See text and Fig.\,\ref{fig:DUNE sensitivity} for more details.
}
\label{tab:DUNE}
\vspace{5pt}
\begin{tabular}{c|c|c|c|ccc}
\toprule
{\multirow{2}{*}{int.}} & \multirow{2}{*}{DUNE} & \multirow{2}{*}{CHARM II~\cite{Chen:2021uuw}} & \multirow{2}{*}{LEP~\cite{Fox:2011fx}}  & \multicolumn{3}{c}{future lepton colliders~\cite{Ge:2023wye}} \\
\cline{5-7}
& & & & CEPC & ILC & CLIC \\
\midrule
S/P & 560 & 348 & 440 & 2.0$\times 10^3$ & 2.6$\times 10^3$ & 3.2$\times 10^4$ \\ \hline
V & \multirow{2}{*}{650} & \multirow{2}{*}{493} & 470 & \multirow{2}{*}{2.2$\times 10^3$} & \multirow{2}{*}{2.8$\times 10^3$} & \multirow{2}{*}{5.0$\times 10^4$} \\ 
A & & & 480 & & & \\ \hline
T & 1070 & 779 & $-$ & 1.8$\times 10^3$ & 2.5$\times 10^3$ & 5.3$\times 10^4$ \\ 
\bottomrule
\end{tabular}
\end{table}

There are some studies of the sensitivities of DUNE ND in simplified models with light mediators~\cite{Candela:2024ljb} (see also Ref.~\cite{Chakraborty:2021apc} for the general neutrino-electron interactions with a light mediator). Our estimates in the framework of effective dimension-6 couplings in \geqn{lagrangian} correspond to the heavy mediator limit, analogous to the Fermi theory description of the weak interaction. While the analysis of Ref.~\cite{Candela:2024ljb} is performed in the light mediator framework with an explicit propagator dependence, it is nevertheless instructive to compare the order-of-magnitude sensitivities of these two approaches. Formally, the heavy mediator EFT can be related to simplified models through the correspondence,
\begin{equation}
  \frac{1}{\Lambda^2}
\sim
  \frac{g_{\rm eff}^2}{m_{\rm med}^2},
\end{equation}
with $g_{\rm eff}$ being the effective coupling involved. This correspondence is valid only in the limit of $m_{\rm med} \gg |\bm q|$.
At the DUNE experiment, the typical momentum transfer in neutrino-electron scattering is
$|\bm q| \sim \mathcal O(10\!-\!50)\,\mathrm{MeV}$. The projected sensitivities for light mediators over a range of benchmark parameters are obtained in Ref.~\cite{Candela:2024ljb}. Taking, for instance, the benchmark scenario with $m_\chi = 0.1\, m_{\rm med}$ (with $m_{\rm med}$ being the mediator mass), we find that the sensitivities to the cutoff scales $\Lambda_{i,e}$ obtained in this work are consistent
with the corresponding sensitivities of the effective coupling $g_{\rm eff}$ at the DUNE ND in Ref.~\cite{Candela:2024ljb}.

\subsection{CHARM II and LEP limits and prospects at future lepton colliders}
\label{sec:limits:LEP}

The cutoff scales $\Lambda_{i,e}$ are constrained by the $\nu(\bar\nu)$-$e^-$ scattering at the CHARM II experiment. 
Following the analysis of Ref.~\cite{Chen:2021uuw}, the existing CHARM II data yield the following 90\%\,C.L. lower bounds in the limit $m_\chi \to 0$: $\Lambda_{S(P),e}\gtrsim 348\,\mathrm{GeV}$, $\Lambda_{V(A),e}\gtrsim 493\,\mathrm{GeV}$ and $\Lambda_{T,e}\gtrsim 779\,\mathrm{GeV}$.
These CHARM II bounds are shown as the blue curves in \gfig{fig:DUNE sensitivity} and are summarized in \gtab{tab:DUNE}. 

The effective couplings of neutrinos and $\chi$ with electrons are also constrained by the existing LEP data. The operators of electrons and DM $\chi$
\begin{equation}
\label{eqn:operator:DM}
{\cal O}_i = (\bar\chi \Gamma^i \chi ) (\bar{e} \Gamma_i e)
\end{equation}
can induce mono-$\gamma$ signals at high-energy $e^+ e^-$ colliders. The existing LEP data have excluded the cutoff scales $\Lambda_{i,e} \lesssim 440$\,GeV, 470\,GeV and 480\,GeV for the Lorentz structures of scalar, vector and axial-vector couplings~\cite{Fox:2011fx}. Recast onto the couplings in Eq.~(\ref{lagrangian}), the limits are expected to be the same, as the mass $m_\chi$ is too much lower than the energy scale of LEP. The LEP limits are shown as the shaded regions below the orange dashed lines in \gfig{fig:DUNE sensitivity} and are also collected in \gtab{tab:DUNE}. 

As clearly seen in \gfig{fig:DUNE sensitivity} and \gtab{tab:DUNE}, the DUNE ND can probe large cutoff scales $\Lambda_{i,e}$ beyond the CHARM II and LEP limits for all the five Lorentz structures, for dark particle mass $m_\chi $ up to roughly 50\,MeV. 
This is primarily driven by the significantly higher neutrino flux at DUNE with respect to these older experiments: the statistical advantage from the intense LBNF beam compensates for the lower center-of-mass energy with respect to LEP, and the much larger event sample compared to CHARM II. This demonstrates that modern high-intensity neutrino beam experiments such as DUNE can provide unique and competitive sensitivity to the dark sector physics, largely complementary to the lepton collider searches.

The prospects for probing $\Lambda_{i,e}$ can be significantly improved at the future
high-energy lepton colliders, as detailed in Ref.~\cite{Ge:2023wye} (see also Ref.~\cite{Kundu:2021cmo} for studies of the operator (\ref{eqn:operator:DM}) at future lepton colliders). Early studies of the mono-$\gamma$ channel at the ILC can be found in Ref.~\cite{Chae:2012bq}, which estimates a reach of $\Lambda \sim (1 \sim 1.2)$\,TeV at $\sqrt{s}=250$\,GeV and $\Lambda \sim (3 \sim 4$)\,TeV at $\sqrt{s}=1$\,TeV with polarized beams. Related model-independent analyses of radiative dark particle pair production, $e^+e^- \to \chi\bar\chi\gamma$, were also performed in Refs.~\cite{Bartels:2012ex,Dreiner:2012xm}.
At the future high-energy $e^+ e^-$ colliders, the most sensitive probes of the effective couplings in \geqn{lagrangian} involving electrons are in the channels with final states of mono-$\gamma$ and $e^+ e^- + \slashed{E}_T$. The corresponding sensitivities of $\Lambda_{i,e}$ depend largely on the center-of-mass energy $\sqrt{s}$ and the integrated luminosity. Combining both channels, it is found that $\Lambda_{i,e}$ can be probed up to $\sim 2$\,TeV at the CEPC with $\sqrt{s} = 240$\,GeV and a luminosity of $5.6$\,ab$^{-1}$. At the ILC with $\sqrt{s} = 500$\,GeV and 4\,ab$^{-1}$ and the CLIC with $\sqrt{s} = 3$\,TeV and 5\,ab$^{-1}$, the sensitivities of $\Lambda_{i,e}$ can be improved up to $\sim 2.5$\,TeV and $\sim 30$\,TeV. Conservatively, all the beams are assumed to be unpolarized. These CEPC, ILC and CLIC prospects are presented in \gfig{fig:DUNE sensitivity} as the solid, dashed and dot-dashed green lines, respectively. The corresponding specific values of $\Lambda_{i,e}$ are collected in \gtab{tab:DUNE}.


\subsection{Other weak constraints}
\label{sec:limits:EnuES:weaker}

We now list some other constraints that are relatively weak or not applicable to our framework.
\begin{itemize}
\item \textbf{Dark particle searches at $B$-factories}. The sub-GeV dark particles can also be searched for at the $B$-factories via the mono-photon process of $e^+e^- \to \gamma \chi \bar\chi$. However, with an integrated luminosity of 50\,ab$^{-1}$ at Belle II, the cutoff scales $\Lambda_{i,e}$ can only be probed up to roughly 300\,GeV, significantly lower than the current LEP limits~\cite{Liang:2021kgw,Liang:2023lwh} and the prospects at future high-energy lepton colliders~\cite{Ge:2023wye,Kundu:2021cmo,Bartels:2012ex,Chae:2012bq,Dreiner:2012xm}.    
    
\item \textbf{Dark particle searches at beam-dump experiments}. 
Given the operator in \geqn{eqn:operator:DM}, dark particles can be pairly produced at the electron beam dump experiments via the process of $e^- + \gamma \to e^- + \chi + \bar\chi$, where the photon is emitted from the target nuclei. The constraints from the existing NA64 data are rather weak, with $\Lambda_{i} \gtrsim 10$\,GeV~\cite{Liang:2021kgw}. 
    
\item \textbf{Dark particle with electromagnetic form factors}. There have been studies of dark particles with electromagnetic form factors, which will induce the pair production of dark particle $\chi$ at the electron beam experiments such as BaBar, NA64 and mQ, mediated by a virtual photon~\cite{Chu:2018qrm}. Re-interpreting onto the cutoff scales $\Lambda_{i,e}$ is constrained up to the scale of 200\,GeV. Future fixed-target experiments such as LDMX and BDX can probe the magnetic dipole moment $\mu_\chi \sim 10^{-6}\,\mu_B$, which corresponds to $\Lambda \gtrsim \mathcal{O}(1\,\text{TeV})$.

\item \textbf{Low-energy neutrino experiments}. The dark particle $\chi$ can be produced at the low-energy neutrino experiments TEXONO, and Borexino. However, these experiments are only sensitive to the mass range of $m_\chi \lesssim 1$\,MeV~\cite{Chen:2021uuw}. 

\item 
\textbf{Neutrino and dark trident processes}. The neutrino trident process can be induced by a (light) mediator that couples to neutrinos and other SM particles. The effective parton-level process can be written as $\nu \gamma \to \nu f \bar{f}$, with the photon emitted from the target. With a gauge boson mediator $Z'$, such a process interferes constructively with the SM contribution, and the effective cutoff scale is excluded up to ${\cal O} (500\,{\rm GeV})$ by the CCFR experiment~\cite{CCFR:1991lpl,Altmannshofer:2014pba}. Replacing the neutrino in the final state by the dark fermion $\chi$, i.e. $\nu \gamma \to \chi f \bar{f}$, this process will not interfere any more with the SM one, and the corresponding limits on the cutoff scales are expected to be significantly weaker. 

The dark trident process is very similar to the neutrino one, with the parton-level process of $\chi A \to \chi f \bar{f}$ (the gauge boson $A$ here can be a photon or a beyond SM mediator such as a dark photon). The expected sensitivities at SBND, MicroBooNE, ICARUS, DUNE ND and DarkQuest have been estimated in Refs.~\cite{deGouvea:2018cfv,Dutta:2024nhg}, for the couplings via a light mediator. With a neutrino in the final state, i.e. $\chi A \to \nu f \bar{f}$, the corresponding sensitivities are expected to be similar.

\item \textbf{Light mediator portal dark particle searches}. Dark particles can couple to electrons via a light mediator~\cite{Batell:2021blf,BaBar:2017tiz,Abdullahi:2023tyk,Asai:2023dzs,DeRomeri:2019kic,Batell:2009di,Izaguirre:2014dua,deNiverville:2011it,Balan:2024cmq,Dutta:2020vop,Catena:2023use,Mongillo:2023hbs,Morrissey:2014yma,Voronchikhin:2023znz,Asai:2022zxw,Asai:2023mzl}. However, as discussed in \gsec{sec:weaker:cevns}, the constraints can not be directly recast onto the cutoff scales in \geqn{lagrangian}.

\item \textbf{Four-body meson and tauon decays}. The couplings of neutrinos and dark particle $\chi$ with electron will induce some exotic charged lepton and meson decays, e.g. the four-body tauon decay $\tau^- \to \pi^- e^+ e^- \chi$ and the meson decay $\pi^+,\, K^+ \to \ell^+ e^+ e^- \chi$, with the $e^+ e^-$ pair emitted from the neutrino line. However, these channels is highly suppressed by the phase space, and the limits from experimental measurements in the channels $\tau^- \to \pi^- e^+e^- \nu_\tau$~\cite{Belle:2019bpr}, $\pi^+ \to e^+ e^+ e^- \nu_e$~\cite{SINDRUM:1989qan} and $K^+ \to \ell^+ e^+ e^- \nu_\ell$ ($\ell = e,\,\mu$)~\cite{Poblaguev:2002ug}   
are expected to be rather weak.

Similarly, the four-body meson decays $\pi^-,\, K^- \to e^- \nu (\nu \bar\chi,\, \chi \bar\nu)$ are also possible, with the $(\nu \bar\chi)$ or $(\chi \bar\nu)$ pair connected to the final-state electron line. In the limit of $m_\chi \to 0$, the corresponding signals look quite similar as the decays $\pi^-,\, K^- \to e^- \nu \nu \bar\nu$, which can be induced by neutrino self-interactions~\cite{Bilenky:1999dn,Bardin:1970wq,Kelly:2019wow}. However, the corresponding limits from the data of $\pi^+ \to e^+ \nu_e \nu \bar\nu$~\cite{PIENU:2020las} and $K^+ \to e^+ \nu_e \nu \bar\nu$~\cite{Heintze:1977kk} are also very weak (cf. the relevant discussions in Ref.~\cite{Dev:2025tdv}).     
    
\item \textbf{Five-body muon and tauon decays}. The operator in \geqn{lagrangian} will induce the five-body muon decays $\mu^- \to e^- e^+e^- (\nu_\mu \bar\chi,\, \chi \bar\nu_e)$ and tauon decays $\tau^- \to \ell^- e^+e^- (\nu \bar\chi,\, \chi \bar\nu)$ (with $\ell = e,\,\mu$), with the $e^+ e^-$ pair emitted from the (anti)neutrino line. 
These channels are constrained by measurements of the SM processes $\mu^- \to e^- \nu_\mu \bar\nu_e e^+e^-$~\cite{SINDRUM:1985vbg} and $\tau^- \to \ell^- e^+ e^- \nu \bar\nu$~\cite{CLEO:1995azm}. With extra neutrinos and the dark particle $\chi$ emitting from the electron line in the final state, there are also the five-body charged lepton decays $\mu,\,\tau \to e \nu \bar\nu (\nu \bar\chi,\, \chi\bar\nu)$, which look quite similar to the SM processes $\mu,\,\tau \to e \nu \bar\nu$ and the four-body decays $\mu,\, \tau \to e \nu \bar\nu \phi$ with $\phi$ a neutrinophilic scalar~\cite{Bickendorf:2022buy,deGouvea:2019qaz,Brdar:2020nbj}.  However, the corresponding limits in these channels are expected to be very weak due to phase space suppression. 

\item \textbf{Four-body $W$ boson decays}. There might be the rare decay of $W \to e \nu (\nu \bar\chi,\, \chi \bar\nu)$ induced by the couplings in \geqn{lagrangian}, with (anti)neutrino and $\chi(\bar\chi)$ emitted from the electron line. These decay channels are quite similar to $W \to e \nu \nu \bar\nu$, which can be induced by neutrino self-interactions~\cite{Bilenky:1992xn,Bilenky:1999dn}. However, suppressed again by the four-body phase space, the constraints from the $W$ data are rather weak~\cite{ParticleDataGroup:2024cfk}. 

\item \textbf{Radiative $Z$, $W$ and meson decays}. 
At the 1-loop order, the couplings of neutrinos and dark particle with electrons in \geqn{lagrangian} induce the exotic radiative decays $Z \to \nu \bar\chi,\,\chi\bar\nu$~\cite{Davidson:2003ha}.
The precision measurements of the invisible partial width of $Z$ boson set constraints on the cutoff scales $\Lambda_{i,e}\gtrsim {\cal O}(100\,{\rm GeV})$. 
The precision measurements of $W \to e \nu$ can also be used to constrain the decay channel of $W \to e\chi$, which is induced by the couplings of neutrinos and dark particle $\chi$ with electrons~\cite{Davidson:2003ha}, but the corresponding precision are to some extent lower than that from $Z$ data, which leads to weaker constraints on the cutoff scales~\cite{ParticleDataGroup:2024cfk}. The couplings in \geqn{lagrangian} could also induce the radiative exotic meson decay ${\sf M}^- \to e^- \bar\chi$, which, however, is highly suppressed by the loop factor and electron mass (cf. the 1-loop corrections to the meson decays ${\sf M}^- \to e^- \bar\nu$ in Ref.~\cite{Dev:2024ygx}).

\item \textbf{DM direct detection}. In the case of $m_\chi < 1$\,MeV, $\chi$ could play as decaying DM, and induce electron recoil signals via the process of $\chi + e^- \to e^- + \nu$ in the DM direct detection experiments~\cite{Dror:2020czw,Ge:2022ius}. However, the corresponding stringent limits from DM direct detection experiments \cite{PandaX:2022ood,EXO-200:2022adi,CDEX:2024bum} are not applicable to the case of $m_\chi >1$\,MeV, as a result of the rapid decay of $\chi$.
    




\item \textbf{Supernova observations}. The dark particle can be produced in the supernova core via the processes $\nu + e \to \chi + e$ and $e^+ + e^- \to \chi(\bar\chi) + \bar\nu(\nu)$. Therefore, the effective couplings in \geqn{lagrangian} for electrons are constrained by the supernova observations, e.g. from the cooling of SN1987A~\cite{Lin:2025mez,Manzari:2025jbc}. However, for $m_\chi > 2m_e$, $\chi$ could decay via $\chi \to \nu e^+ e^-$ inside the supernova core (cf. \geqn{eqn:width}) and weaken or invalidate the supernova cooling constraints in some regions of the parameter space. The supernova limits in Refs.~\cite{Lin:2025mez,Manzari:2025jbc} are therefore only for $m_\chi <1$\,MeV, not applicable to the mass range of $m_\chi > 1$\,MeV in this paper.

\end{itemize}

\section{Conclusion}
\label{sec:conclusion}

In this work, we have performed a model-independent analysis of the effective couplings of neutrino, dark particle $\chi$ with nucleons or electrons at some representative neutrino scattering experiments, for all the five possible Lorentz structures of scalar, pseudoscalar, vector, axial-vector and tensor interactions. For the hadronic absorption operators, the current COHERENT and CONUS+ data could exclude the cutoff scales $\Lambda_{i,N}$ up to the scale of few hundreds of GeV, less stringent than the existing LHC data and SN1987A observations (cf. \gfig{fig:CEvNS sensitivity} and \gtab{tab:cevns}). However, for the leptonic absorption operators, the cutoff scales $\Lambda_{i,e}$ for couplings with electrons can be probed up to $\sim 1$\,TeV at the DUNE ND, with the dark particle mass $m_\chi$ up to $\sim 100$\,MeV, depending to some extent on the Lorentz structure. The DUNE ND sensitivities are beyond the current CHARM II and LEP limits (cf. \gfig{fig:DUNE sensitivity} and \gtab{tab:DUNE}), and are largely complementary to the searches at future high-energy lepton colliders.




\section*{Acknowledgments}

The authors would like to thank Xiao-Dong Ma for useful discussions. The authors are also very grateful to the referee for the valuable comments and suggestions. The work of RF and YZ is supported by the National Natural Science Foundation of China (NSFC) under Grant No. 12175039. The work of SFG is supported by the National NSFC under the Grant No. of 12375101 and 12425506. This work is also supported by
State Key Laboratory of Dark Matter Physics and the Fundamental Research Funds for the Central Universities. 


\appendix 

\section{Weak nuclear form factor and spin structure functions}
\label{app:nuclear}

In this appendix we collect the nuclear physics inputs used in \geqn{eq:dSigmadT}, namely the weak nuclear form factor $F_W(|\bm q|^2)$ and the spin structure functions $\tilde S^{\mathcal T}(|\bm q|^2)$ and $\tilde S^{\mathcal L}(|\bm q|^2)$, with the three-momentum transfer $|\bm q|^2 \simeq 2 m_{\mathcal N} T_{\mathcal N}$.


The weak nuclear form factor $F_W(|\bm q|^2)$ accounts for the finite size and nuclear physics effects of the target nucleus. Following Refs.~\cite{Klein:1999qj,DeRomeri:2024hvc}, we adopt the Klein-Nystrand parametrization,
\begin{align}
F_W(|\bm q|^2)=\frac{3\,j_1(|\bm q|R_{\sf A})}{|\bm q|R_{\sf A}\left(1+a_k^2|\bm q|^2\right)} \,,
\end{align}
where $j_1(x)=\sin x/x^2-\cos x/x$ is the spherical Bessel function of order one, $R_{\sf A} = 1.23\,\sf A^{1/3}~{\rm fm}$ is the effective nuclear radius, 
and $a_k=0.7~{\rm fm}$ smears the nuclear
surface.




Note that $\tilde{S}^{\mathcal{T}} (|\bm q|^2)$ and $\tilde{S}^{\mathcal{L}} (|\bm q|^2)$ are the transverse and longitudinal nuclear spin structure functions, respectively, which are relevant to the axial-vector and tensor interactions.  
We follow the standard multipole decomposition of the nuclear current and encode nuclear structure effects in spin structure functions, as commonly adopted in studies of spin-dependent neutrino-nucleus scattering~\cite{DeRomeri:2024hvc,Hoferichter:2020osn}. It is convenient to define 
\begin{align}
S^\kappa_X(|\bm q|^2)
 \ = \ (g_X^0)^2\,S_{00}^\kappa(|\bm q|^2)
+ g_X^0 g_X^1\,S_{01}^\kappa(|\bm q|^2)
+ (g_X^1)^2\,S_{11}^\kappa(|\bm q|^2)\,,
\label{eq:SpinDecomp}
\end{align}
where $X=A,\,T$ stands for the interaction type, $\kappa=\mathcal T,\mathcal L$ denotes the transverse ($\mathcal T$) or longitudinal ($\mathcal L$) component, and $S_{ij}^\kappa$ are the nuclear spin response functions depending on the target nucleus~\cite{Hoferichter:2020osn,Kosmas:2021zve}. 
The couplings $g_X^0$ and $g_X^1$ are the isoscalar and isovector
combinations of the proton and neutron couplings:
\begin{align}
g_X^0 = \frac{g_X^p+g_X^n}{2}\,,
\qquad
g_X^1 = \frac{g_X^p-g_X^n}{2}\,,
\label{eq:isoscalar_isovector}
\end{align}
with $g_X^{p,n}$ the effective proton and neutron couplings, respectively, for the corresponding current $X$.

\section{Nucleon form factors}
\label{app:nucleon}

To quantitatively establish the correspondence between the quark level and nucleon level cutoff scales in \gsec{sec:LHC:cevns}, we adopt the nucleon form factors in Refs.~\cite{Bishara:2017pfq,Ma:2024tkt}. For the low-energy scattering processes of interest, we take the leading contribution in the limit of vanishing momentum transfer, i.e. $q^2 \simeq 0$. Assuming flavor universal couplings to the light quarks $q=u,d,s$ and isospin conservation, the conversion factors ${\cal F}_i$ introduced above can be written as
\begin{equation}
{\cal F}_i \;=\; \sum_{q=u,d,s} {\cal F}_i^{q/N}\,,
\end{equation}
where ${\cal F}_i^{q/N}$ are defined through the nucleon matrix elements of the corresponding quark bilinears.

Using the conventions and numerical inputs of Ref.~\cite{Ma:2024tkt}, the conversion factors for the five Lorentz structures are given by
\begin{subequations}
\begin{align}
{\cal F}_S &\equiv \sum_{q=u,d,s}\frac{\sigma_q^N}{m_q}\simeq 3.90 \,, \\
{\cal F}_P &\equiv \sum_{q=u,d,s}\frac{F_P^{q/N}(0)}{m_q}\simeq 6.43\,, \\
{\cal F}_V &\equiv \sum_{q=u,d,s}F_1^{q/N}(0)=3\,, \\
{\cal F}_A &\equiv \sum_{q=u,d,s}\Delta q^N\simeq 0.49\,, \\
{\cal F}_T &\equiv \sum_{q=u,d,s} g_T^{q/N}\simeq 0.59\,,
\end{align}
\end{subequations}
where $\sigma_q^N$ denotes the nucleon sigma terms, $\Delta q^N$ is the axial-vector spin fractions, and $g_T^{q/N}$ is the tensor charges. All numerical values are taken from Ref.~\cite{Ma:2024tkt} in the $\overline{\mathrm{MS}}$ scheme at $\mu=2\,\mathrm{GeV}$.



\bibliographystyle{utphysGe}
\bibliography{ref}

\end{document}